\pdfoutput=1

\documentclass[11pt]{article}

\usepackage{acl}

\usepackage{times}
\usepackage{latexsym}
\usepackage{multirow}
\usepackage{graphicx}
\usepackage{amsmath}
\usepackage{amssymb}
\usepackage{booktabs}
\usepackage{algorithm2e}

\usepackage{subcaption}
\usepackage{caption}
\usepackage{bbding}
\usepackage{colortbl}
\usepackage{tcolorbox}
\usepackage[T1]{fontenc}
\usepackage{xcolor}         
\usepackage{color}
\usepackage{colortbl,array}
\usepackage[rgb]{xcolor}
\usepackage{tikz}
\usetikzlibrary{tikzmark}
\usepackage{arydshln}
\makeatletter

\usepackage[normalem]{ulem}
\useunder{\uline}{\ul}{}
\newcommand*\myfontsize{%
  \@setfontsize\myfontsize{8}{8}%
}

\makeatother

\definecolor{myred}{rgb}{0.7, 0.3, 0.0}
\definecolor{myblue}{RGB}{235,245,250}
\definecolor{mygreen}{HTML}{056b34}
\definecolor{myorange}{HTML}{ff8800}
\definecolor{mypurple}{HTML}{8400ff}
\definecolor{mypink}{HTML}{f7acb9}

\definecolor{green}{RGB}{0,120,0}
\definecolor{deepblue}{RGB}{0,0,255}
\definecolor{blue}{RGB}{0,179,241}
\definecolor{orange}{RGB}{200,110,0}
\definecolor{purple}{RGB}{120,0,160}

\newcommand{\ie}{\textit{i.e.}}
\newcommand{\eg}{\textit{e.g.}}
\usepackage{algorithm}
\usepackage{algorithmic}

\usepackage[T1]{fontenc}

\usepackage[utf8]{inputenc}

\usepackage{microtype}
\usepackage{bbold}

\usepackage{inconsolata}

\title{Training Documents Reranker with Search Rubrics for Deep Research Agent}

\author{Wenhan Liu$^1$, Yu Lu$^2$, Qiaolin Xia$^2$, Hui Xu$^2$, Tong Zhao$^1$, Jian Xi$^2$,Yutao Zhu$^1$ \\ 
\textbf{Haijin Liang}$^2$, \textbf{Haibo Shi}$^2$, \textbf{Hao Wang}$^2$\thanks{Corresponding author.} \and \textbf{Zhicheng Dou}$^{1}$\thanks{Corresponding author.} \\
$^1$Gaoling School of Artificial Intelligence, Renmin University of China \\
$^2$Tencent, Beijing, China \\
\texttt{lwh@ruc.edu.cn, dou@ruc.edu.cn}
}

\begin{document}
\maketitle

\begin{abstract}
Retrieval systems help deep research agents generate high-quality answers by providing relevant documents. However, existing retrievers typically select documents through relevance matching, while individually well-matched top-$k$ documents may not form a \textit{set} that satisfies the complex information needs of an agent query (\eg, diverse, concise and authoritative documents). In this paper, we propose search-oriented rubrics that \textit{explicitly} define the requirements that high-quality document sets should satisfy for each agent query. Our search rubrics are organized into a hierarchical structure and synthesized using a powerful LLM. Based on these search rubrics, we further train a document reranker \textbf{RubricRanker} to select a high-quality subset from retrieved documents. We design a two-stage training framework that consists of rubrics-guided supervised fine-tuning and rubric-based reinforcement learning. Extensive experiments demonstrate that RubricRanker outperforms the strongest baseline by 2.6 points on four deep research benchmarks and generalizes well to five RAG benchmarks.
\end{abstract}

\section{Introduction}\label{sec:intro}
Deep research agents aim to produce in-depth and well-supported answers to complex information-seeking tasks by planning, searching, and synthesizing evidence from diverse sources~\cite{deepresearch_survey,deepresearch_survey1,Dr-tulu,tongyideepresearch}. Unlike conventional single-turn retrieval settings, a deep research agent interacts with search system over multiple steps: it reasons about the current information need, issues an agent query, observes retrieved documents, and decides whether further search is needed before producing a long-form answer. As a result, the ability of search system (\ie, document retrievers and rerankers) directly affects the agent's ability to gather sufficient evidence and generate a reliable final response.

\begin{figure}[t]
	\centering
	\includegraphics[width=1\linewidth]{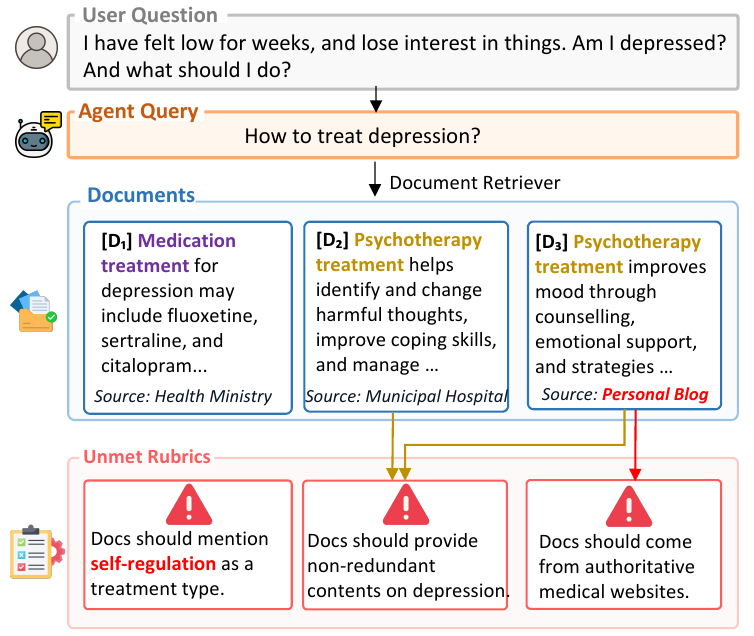}
	\caption{Document-set selection based on single-document relevance matching cannot satisfy the complex information needs of agent queries, such as coverage of diverse aspects, conciseness, and authority.}\label{fig:intro}
\end{figure}

Existing document retrievers~\cite{bge,e5} and rerankers~\cite{rankt5,monot5,rankzephyr} typically return the top-$k$ documents based on relevance matching. Their training supervision is usually based on relevance labels, which measure the relevance of \textbf{single document}. However, it cannot ensure that the returned \textbf{document set} satisfies the complex information needs of an agent query. As illustrated in Figure~\ref{fig:intro}, to answer a help-seeking question from a patient who may have depression, the agent issues the open-ended query ``how to treat depression''. Although all retrieved documents are relevant to depression, the resulting set remains suboptimal for several reasons: (1) it misses an important type of depression treatment, self-regulation; (2) D2 and D3 contain redundant information about psychotherapy, wasting the agent's context budget; and (3) D3 comes from a non-authoritative source, which may lead the agent to generate unreliable medical advice. 

To address this limitation, we propose search-oriented rubrics that specify the requirements that a high-quality document set should satisfy for each query. Our search rubrics serve as query-specific supervision signals to train our document reranker, \textbf{RubricRanker}, which selects a high-quality subset from retrieved documents. To construct search rubrics, we first define hierarchical meta rubrics that cover relevance, conciseness, and consistency at the set level, as well as source authority and timeliness at the document level. We then use a powerful LLM GPT-5.1 to construct a high-quality reference answer for each query to identify the key aspects, facts, and constraints that the selected documents should support. Finally, based on the query and its reference answer, we use GPT-5.1 to expand the meta rubrics into query-specific rubrics with corresponding importance weights.

Based on the constructed search rubrics, we propose a two-stage training framework, consisting of rubrics-guided supervised fine-tuning (SFT) and rubric-based reinforcement learning (RL), to train our RubricRanker. In the SFT stage, we apply powerful teacher model to select the documents that best satisfy the query-specific rubrics and provide silver labels for cold-start training. In the RL stage, we use the search rubrics to evaluate the document sets selected by the policy model. We design a hierarchical reward aggregation method that combines the set-level and document-level rubric scores into a rubric RL reward. At inference time, RubricRanker directly selects a high-quality subset from the retrieved candidates given the query, \textbf{without requiring the search rubrics}.

Extensive experiments demonstrate that RubricRanker achieves superior performance on deep research benchmarks and generalizes well to RAG benchmarks. Further experiments show that RubricRanker significantly reduces the number of agent search calls, enabling agents to generate better answers with lower latency.

The contributions of this paper are summarized as follows:
\begin{itemize}
    \item We propose search-oriented rubrics that define the properties of high-quality document sets from both set-level and document-level perspectives.
    \item We propose a two-stage training framework that combines rubrics-guided SFT with rubric-based RL to train a document reranker for both deep research and RAG tasks.
    \item Extensive experiments demonstrate that RubricRanker achieves strong performance on both deep research and RAG benchmarks.
\end{itemize}

\section{Related Work}

\paragraph{RAG and Deep Research}
Retrieval-augmented generation (RAG) enables LLMs to access external knowledge beyond their parameters~\cite{rag_survey}. However, single-turn retrieval is often insufficient for complex information-seeking tasks that require iterative evidence gathering. This has motivated agentic search systems that interleave reasoning with retrieval~\cite{react}, as well as deep research agents that combine reasoning models with web search to solve open-ended queries~\cite{deepresearch_survey1,tongyideepresearch}. In these agents, retrieval quality affects not only the current observation but also subsequent reasoning and search decisions, so errors can accumulate across multiple steps. We therefore focus on improving the document sets returned for each agent-issued query.

\paragraph{Document Ranking}
Document ranking orders candidate documents by query--document relevance, modeled through dense representations~\cite{bge,e5}, interaction-based matching~\cite{monot5}, or LLM-based ranking~\cite{rankgpt,prp,fullrank,reasonrank,coranking,demorank,sumrank}. Although these methods perform well on standard IR benchmarks, they mainly optimize relevance rankings of individual documents. With the development of RAG~\cite{rag_survey} and deep research agents~\cite{deepresearch_survey}, recent studies have instead optimized rerankers for downstream LLM generation. Rank4Gen~\cite{rank4gen} trains rerankers with supervision derived from answer-generation quality, while \citet{setr} and \citet{zhang2025distilling} distill document-selection preferences from strong teacher models. However, such signals are difficult to obtain for open-ended queries whose answers are hard to verify, and answer-level preferences provide limited guidance about which properties of the selected document set should be improved. They may therefore miss explicit requirements such as conciseness, consistency, and source quality.

Rubrics provide structured criteria for evaluating open-ended outputs~\cite{rubrics_survey} and interpretable reward feedback~\cite{rar,Rubicon,advancedif,Dr-tulu}. For example, RAR~\cite{rar} uses instance-specific rubrics as rewards for reinforcement learning on open-ended medical and scientific tasks. Inspired by these works, we propose search-oriented rubrics that explicitly define fine-grained document-set quality and use them to train rerankers for agent generation.

\section{Preliminary}
In this section, we describe how document reranking is incorporated into deep research and retrieval-augmented generation (RAG).

\subsection{Document Reranking in Deep Research}
Deep research agents answer complex information-seeking questions through iterative reasoning and external information acquisition. We adopt the ReAct paradigm~\cite{react}, which is widely used in deep research agents~\cite{Dr-tulu,tongyideepresearch}. At each step, the agent maintains an interaction history, produces a \textit{Thought} and an \textit{Action}, and receives an \textit{Observation} from the environment. Each intermediate action corresponds to a tool call, such as web search. The observation is appended to the agent context and influences both subsequent reasoning and whether additional evidence should be collected. The agent repeats this process until it has sufficient information to produce a long-form answer.


We insert the document reranker after the agent invokes the search tool. Specifically, for a search action, the agent issues a query $q_t$, which we refer to as a \textbf{sub-query} to distinguish it from the original user question. The search API returns a list of web documents $\mathcal{D}_t = \{d_1, \ldots, d_n\}$. Given the sub-query $q_t$ and document list $\mathcal{D}_t$, the reranker ranks the documents and selects a high-quality subset $\mathcal{S}_t \subseteq \mathcal{D}_t$, which is then provided to the deep research agent as the observation for this search step.

\subsection{Document Reranking in RAG}
RAG follows a simpler single-turn retrieval-and-generation paradigm. Given an input user question $q$, a retriever first searches an external corpus and returns a candidate document list $\mathcal{D} = \{d_1, \ldots, d_n\}$. The document reranker reranks these candidates to select a high-quality document subset $\mathcal{S} \subseteq \mathcal{D}$ as evidence for answer generation. Unlike in deep research, retrieval is performed for the original user question $q$ rather than for agent-generated sub-queries.

\section{Methodology}
\begin{figure*}[!tb]
  \centering
  \includegraphics[width=.95\linewidth]{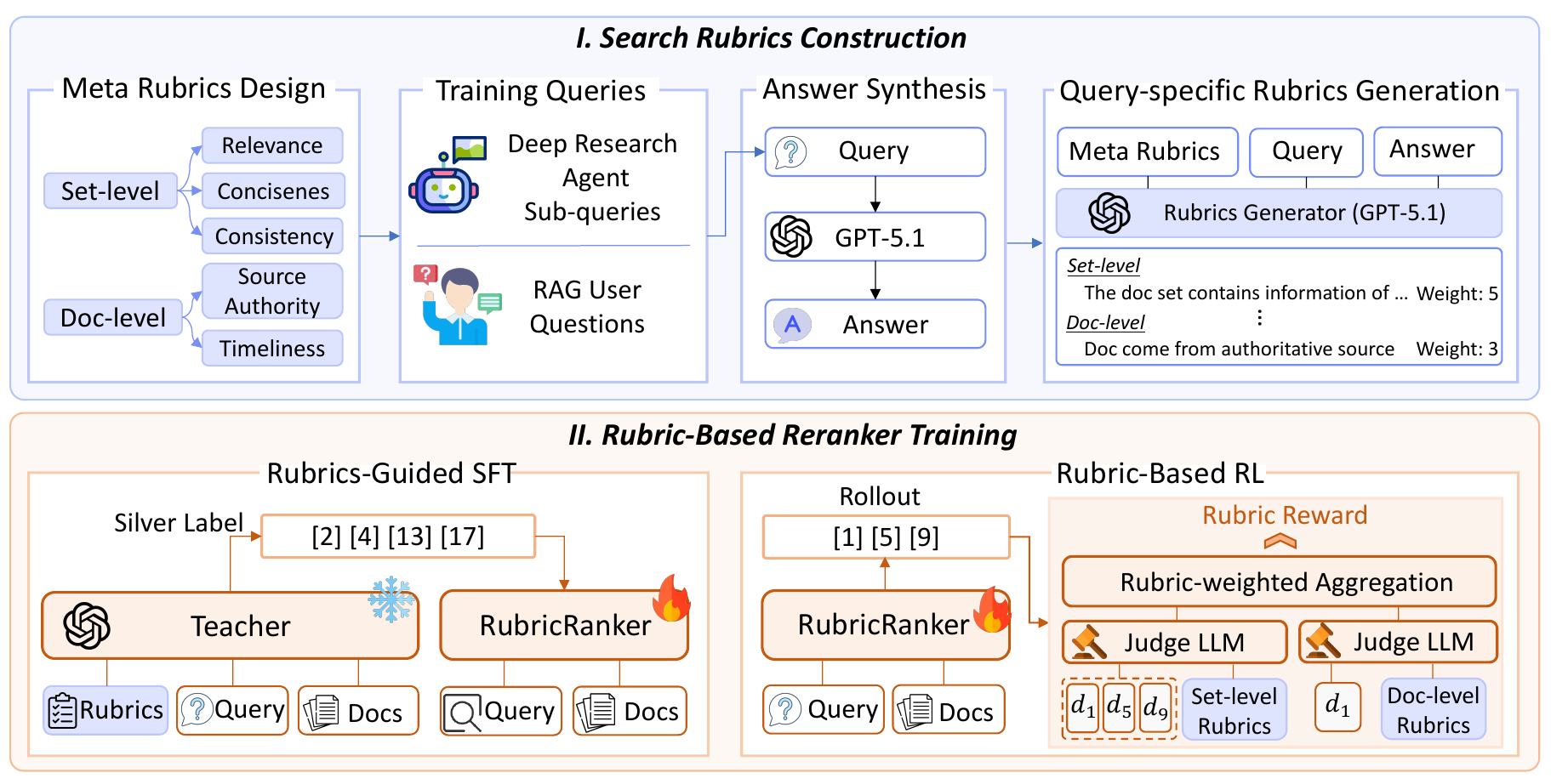}
  \caption{An overview of our framework, which constructs query-specific search rubrics and uses them to train a document reranker through rubrics-guided SFT and rubric-based RL.}\label{fig:model}
\end{figure*}

This section presents the design of RubricRanker. The key idea is to express the characteristics of high-quality document sets as explicit search rubrics and then use these rubrics to train the reranker. Specifically, our framework consists of two parts: constructing query-specific search rubrics and training the reranker with rubric-based supervision. Figure~\ref{fig:model} illustrates the overall framework.
\subsection{Search Rubrics Construction} \label{subsec:rubric_construct}
We design query-specific search rubrics that capture the properties a selected document set should satisfy at both the set and document levels. First, we design hierarchical meta rubrics as a general framework for query-specific rubric generation. We then collect training queries from both deep research and RAG scenarios and use a strong LLM to synthesize high-quality answers to these queries. Finally, we use a strong LLM to generate query-specific rubrics based on the meta rubrics, each query, and its synthesized answer.

\subsubsection{Meta Rubrics Design.}
A widely used approach to rubric generation is to prompt LLMs to produce rubrics directly~\cite{Rubicon,rar}. However, direct prompting may produce rubrics with insufficient coverage, repetitive content, or conflated evaluation dimensions~\cite{RRD}. We therefore propose meta rubrics as a fixed framework for query-specific rubric generation. Our meta rubrics adopt a \emph{hierarchical structure} that covers both set-level properties of the selected document set and document-level properties of each document.

\paragraph{Set-level rubrics.}
Set-level rubrics evaluate the overall quality of a document set. We consider three dimensions: (1) \textbf{Relevance}, which measures whether the document set covers the key information needs and diverse answer aspects required by the query; (2) \textbf{Conciseness}, which measures whether the set contains little repetitive, redundant, or irrelevant content; and (3) \textbf{Consistency}, which measures whether key facts, claims, and conclusions across documents are mutually compatible rather than contradictory.

\paragraph{Document-level rubrics.}
Document-level rubrics evaluate the quality of a single document. We consider two dimensions: (1) \textbf{Source authority}, which measures whether a document provides \textit{relevant} evidence from an \textit{authoritative and reliable} source, especially in high-stakes domains such as medicine, law, and policy; and (2) \textbf{Timeliness}, which measures whether a document provides \textit{relevant} evidence that satisfies explicit temporal constraints or needs for up-to-date information. Both dimensions are defined only for relevant documents; that is, an irrelevant document does not satisfy the source authority rubric even if it comes from authoritative sources.

\subsubsection{Training Query Collection.}
In the deep research scenario, RubricRanker reranks documents for sub-queries issued by agents. However, existing open-source datasets generally provide only the original user questions rather than these agent-issued sub-queries. To obtain such sub-queries, we collect open-ended questions from open-source deep research datasets, including OpenScholar~\cite{openscholar}, SearchArena~\cite{searcharena}, GlaiveAI-Reasoning-v1-20M,\footnote{\url{https://huggingface.co/datasets/glaiveai/reasoning-v1-20m}} and WebWalker-Silver~\cite{webwalkerqa}. We then use a strong open-source deep research agent, Dr-Tulu-8B~\cite{Dr-tulu}, to generate a full trajectory for each question and extract the sub-queries issued by the agent as training queries. In addition, we collect user questions from RAG datasets, including HotpotQA~\cite{hotpotqa} and NQ~\cite{nq}, whose answers are short and closed-ended. These questions help construct more diverse rubrics and improve the generalization of RubricRanker across different query scenarios.


\subsubsection{Answer Synthesis.}
Generating query-specific rubrics requires identifying which aspects should be covered, which facts are crucial, and what constraints the supporting evidence should satisfy. However, asking an LLM to generate rubrics based solely on the query may not fully capture these information needs, particularly for complex and open-ended agent queries. Therefore, for agent sub-queries in the deep research scenario, we propose to use the powerful GPT-5.1 model (with web search) to synthesize a high-quality answer as an intermediate information-need blueprint. The prompt is shown in Figure~\ref{fig:answer_generation}. By decomposing the query into concrete aspects and claims, the answer provides a richer basis from which the subsequent rubric generator can derive specific and comprehensive evaluation criteria. For user questions in the RAG scenario, we directly use the gold answers provided by the original datasets.


\subsubsection{Query-Specific Rubrics Generation.} 
Based on the training queries and their corresponding reference answers, we use a strong LLM-based rubric generator GPT-5.1 to expand the predefined meta rubrics into query-specific rubrics. The prompts for agent sub-queries in the deep research scenario and user questions in the RAG scenario are shown in Figures~\ref{fig:rubrics_generation_subq-1}--\ref{fig:rubrics_generation_subq-2} and Figures~\ref{fig:rubrics_generation_userq-1}--\ref{fig:rubrics_generation_userq-2}, respectively. Each type of meta rubric may be instantiated as one or more query-specific rubrics, each containing a natural-language description and a weight from 1 to 5, with larger weights indicating more important criteria.

\subsection{Rubric-Based Reranker Training}
Using the constructed query-specific rubrics as supervision signals, we train RubricRanker to select document sets that better satisfy the information needs of each query. The training process consists of two stages. First, we propose rubrics-guided supervised fine-tuning (SFT), in which a teacher model generates silver labels by selecting documents according to the query-specific rubrics, providing a stable cold start for the reranker. Second, to further enhance the reranker's document-set selection ability, we perform rubric-based reinforcement learning (RL), using a hierarchical reward aggregation method that combines set-level and document-level rubric scores to optimize the selected document sets.

\subsubsection{Rubrics-Guided SFT}
Selecting a high-quality document set for a complex query is challenging for the base model, and low-quality rollouts may also hinder subsequent RL training. To address this issue, we use a strong teacher model guided by our constructed query-specific rubrics to produce silver labels for cold-start training.

For each training query, we retrieve a candidate document list with a randomly sampled length ranging from 10 to 40 and use these candidates as training documents. Such variable-length input lists expose RubricRanker to different input sizes during training and enable it to handle different numbers of candidate documents at inference time. For agent sub-queries in the deep research scenario, we retrieve web documents using the Serper Google Search API.\footnote{\url{https://serper.dev/}} For user questions in the RAG scenario, we use BGE-base model to retrieve documents from the Wikipedia corpus.

We then use GPT-5.1 as the teacher model to generate labels. The teacher takes the query-specific rubrics, training query,\footnote{For an agent sub-query, we also provide the teacher and student with the preceding agent reasoning (omitted from the figure for clarity) to clarify the query intent.} and candidate document list as input and outputs a list of selected document IDs (\eg, ``[2] [4] [13] [17]'') that best satisfy the query-specific rubrics. The student reranker is trained on these teacher labels using the query and candidate document list as input, without access to the query-specific rubrics, consistent with the final inference stage.

\subsubsection{Rubric-Based RL}
After cold-start SFT, we further optimize the reranker with reinforcement learning to improve its document-set selection ability. The key motivation is to use rubric weights to compute more accurate rewards for document-set selection and to further refine the reranker through fine-grained comparisons among different rollouts. Given a query and a candidate document list, the reranker outputs a sequence of selected document IDs, which is then parsed as a document set $D$.

Given query-specific rubrics $\mathcal{R}=\{sr_1, \ldots, sr_i\} \cup \{dr_1, \ldots, dr_j\}$, where $sr_i$ and $dr_j$ denote the $i$-th set-level rubric and the $j$-th document-level rubric, respectively, we compute the rubric reward $P^r$ through a hierarchical aggregation with rubric weights:
\begin{equation}
\begin{aligned}
P^r(D) =
\frac{
\sum_i sw_i\!\cdot\!S(sr_i, D)
+ \sum_j dw_j\!\cdot\!F(dr_j, D)
}{
\sum_i sw_i + \sum_j dw_j
},
\end{aligned}
\end{equation}
where $sw_i$ and $dw_j$ are the weights of the set-level rubric $sr_i$ and document-level rubric $dr_j$, respectively. The scoring function $S(\cdot)$ is implemented by a GPT-5.1-based LLM judge, which assigns a normalized score between 0 and 1 according to the given rubric. For set-level rubrics, $S(sr_i, D)$ scores whether the selected documents $D$ collectively satisfy the query requirements. For each document-level rubric $dr_j$, we first use the LLM judge to score each selected document $d$ and then average the scores over the selected set:
\begin{equation}
F(dr_j, D) = \frac{1}{|D|} \sum_{d \in D} S(dr_j, d).
\end{equation}
In addition, we apply a format reward to ensure that the output can be parsed. Specifically, the output must follow a document-ID format such as ``[1] [3] [2]''. If the output format is correct, we parse the selected document IDs as $D$ and use the rubric reward as the final reward; otherwise, the final reward is set to $-1$:
\begin{equation}
P(y) =
\begin{cases}
P^r(D), & \text{if the output format is correct}, \\
-1, & \text{otherwise},
\end{cases}
\end{equation}
Finally, based on the final reward $P(y)$, we apply the GRPO algorithm~\cite{grpo} to optimize RubricRanker. The RL training details are provided in Section~\ref{subsec:rl}.

\section{Experiments}
\subsection{Settings}
\paragraph{Evaluation Benchmarks.}
We evaluate RubricRanker in two scenarios. \textbf{The first scenario is deep research}, for which we use HealthBench~\cite{healthbench}, WebWalkerQA~\cite{webwalkerqa}, DeepResearchBench (DRB)~\cite{deepresearchbench}, and ResearchQA~\cite{researchqa}. Detailed descriptions of these benchmarks are provided in Appendix~\ref{sub:dr_datasets}. For all these benchmarks, responses are expected to be open-ended or long-form and are evaluated by LLM judges using rubrics from the official evaluation protocols. Given the high cost and time overhead of LLM-based evaluation, we randomly sample 100, 200, 100, and 100 queries from HealthBench, WebWalkerQA, DRB, and ResearchQA, respectively, as our test sets.


\textbf{The second scenario is RAG}, for which we use HotpotQA~\cite{hotpotqa}, Bamboogle~\cite{bamboogle}, Natural Questions (NQ)~\cite{nq}, TriviaQA~\cite{triviaqa}, and PopQA~\cite{popqa}. Unlike the deep research benchmarks, these datasets mainly require short, closed-ended answers. Following prior work~\cite{search-r1,agentic-r}, we use exact match (EM) as the evaluation metric.

\paragraph{Baselines.}
We compare RubricRanker with two categories of document rerankers. The first category is \textbf{vanilla rerankers}, which are mainly optimized for query--document relevance to improve the ranking of retrieved documents. This category includes BGE-Reranker-Large\footnote{\url{https://huggingface.co/BAAI/bge-reranker-large}}, MonoT5~\cite{monot5}, RankT5~\cite{rankt5}, RankVicuna~\cite{rankvicuna}, and RankZephyr~\cite{rankzephyr}. The second category is \textbf{generation-oriented rerankers}, which aim to select documents that better support LLM answer generation. This category includes SetR~\cite{setr} and Rank4Gen~\cite{rank4gen}. Detailed descriptions of these baselines are provided in Appendix~\ref{sub:baseline_details}.

\begin{table*}[!t]
\small
\centering
\setlength{\tabcolsep}{2.2mm}{
\begin{tabular}{lcccccc}
\toprule
\textbf{Method} & \textbf{Size} & \textbf{WebWalkerQA} & \textbf{HealthBench} & \textbf{DRB} & \textbf{ResearchQA} & \textbf{Avg.} \\ \midrule
Initial Retrieval (Google Search API) & - & 44.5 & 55.2 & 45.3 & 71.1 & 54.0 \\ \midrule
\multicolumn{7}{l}{\textit{\textbf{Vanilla Rerankers}}} \\
BGE-Reranker-Large & 550M & 52.0 & 58.7 & 46.5 & 71.2 & 57.1 \\
MonoT5 & 3B & 48.0 & 57.8 & {\ul 46.7} & 72.5 & 56.3 \\
RankT5 & 3B & {\ul 53.0} & 57.0 & 46.6 & 72.8 & 57.4 \\
RankVicuna & 7B & 47.0 & {\ul 59.6} & 46.5 & 71.3 & 56.1 \\
RankZephyr & 7B & 50.0 & 58.6 & 46.4 & 71.8 & 56.7 \\ \midrule
\multicolumn{7}{l}{\textit{\textbf{Generation-oriented Rerankers}}} \\
SetR & 8B & 49.0 & 58.7 & 44.8 & {\ul 73.3} & 56.5 \\
Rank4Gen & 8B & 52.0 & 59.2 & 46.6 & 72.0 & {\ul 57.5} \\
\rowcolor{myblue} RubricRanker (Ours) & 8B & \textbf{58.0} & \textbf{61.5} & \textbf{46.8} & \textbf{74.2} & \textbf{60.1} \\ \bottomrule
\end{tabular}}
\caption{Results on deep research benchmarks with Dr-Tulu-8B as the agent. All rerankers rerank the top 30 documents retrieved by the Google Search API, and the final responses are evaluated by an LLM judge using rubrics. The best results are highlighted in \textbf{bold}, and the second-best results are \uline{underlined}.}
\label{tab:main_results_dr}
\end{table*}

\begin{table*}[!t]
\small
\centering
\setlength{\tabcolsep}{2.6mm}{
\begin{tabular}{lccccccc}
\toprule
\textbf{Method} & \textbf{Size} & \textbf{HotpotQA} & \textbf{Bamboogle} & \textbf{NQ} & \textbf{PopQA} & \textbf{TriviaQA} & \textbf{Avg.} \\ \midrule
Initial Retrieval (BGE) & - & 29.8 & 15.2 & 31.0 & 37.2 & 58.4 & 34.3 \\ \midrule
\textit{\textbf{Vanilla Rerankers}} & \multicolumn{1}{l}{\textit{\textbf{}}} & \multicolumn{1}{l}{\textit{\textbf{}}} & \multicolumn{1}{l}{} & \multicolumn{1}{l}{\textit{\textbf{}}} & \multicolumn{1}{l}{\textit{\textbf{}}} & \multicolumn{1}{l}{\textit{\textbf{}}} & \multicolumn{1}{l}{\textit{\textbf{}}} \\
BGE-Reranker-Large & 550M & {\ul 35.3} & 17.6 & 28.0 & 40.4 & 60.8 & 36.4 \\
MonoT5 & 3B & 33.9 & 14.4 & 30.7 & 37.6 & 60.7 & 35.4 \\
RankT5 & 3B & 34.4 & 14.4 & 32.1 & 38.5 & 61.1 & 36.1 \\
RankVicuna & 7B & 33.4 & 14.4 & 30.7 & 38.4 & 59.9 & 35.3 \\
RankZephyr & 7B & 32.8 & 16.0 & 31.9 & 39.5 & 60.6 & 36.1 \\ \midrule
\textit{\textbf{Generation-oriented Rerankers}} & \multicolumn{1}{l}{\textit{\textbf{}}} & \multicolumn{1}{l}{\textit{\textbf{}}} & \multicolumn{1}{l}{} & \multicolumn{1}{l}{\textit{\textbf{}}} & \multicolumn{1}{l}{\textit{\textbf{}}} & \multicolumn{1}{l}{\textit{\textbf{}}} & \multicolumn{1}{l}{\textit{\textbf{}}} \\
SetR & 8B & 34.3 & {\ul 22.8} & 30.3 & 38.5 & 61.2 & 37.4 \\
Rank4Gen & 8B & 35.1 & 20.0 & {\ul 33.2} & {\ul 41.0} & {\ul 61.7} & {\ul 38.2} \\
\rowcolor{myblue} RubricRanker (Ours) & 8B & \textbf{38.0} & \textbf{23.2} & \textbf{34.0} & \textbf{42.2} & \textbf{62.4} & \textbf{40.0} \\ \bottomrule
\end{tabular}}
\caption{Results on RAG benchmarks with Qwen3-8B as the LLM generator. All rerankers rerank the top 30 documents retrieved by BGE, and the generated answers are evaluated using exact match (EM). The best results are highlighted in \textbf{bold}, and the second-best results are \uline{underlined}.}
\label{tab:main_results_rag}
\end{table*}

\paragraph{Implementation Details.}
We use Qwen3-8B as the backbone model for RubricRanker. For deep research benchmarks, we use the Serper Search API for document retrieval. For RAG benchmarks, we use the December 2018 Wikipedia dump~\cite{dpr} as the retrieval corpus and adopt bge-base-en-v1.5\footnote{\url{https://huggingface.co/BAAI/bge-base-en-v1.5}} as the retriever. All document rerankers rank the top 30 retrieved documents and return a document set. For vanilla rerankers that output a full ranking, we use the top five documents. Due to space limitations, we provide further details in Appendix~\ref{app:implementation_details}.

\subsection{Overall Performance} \label{sec:overall}
We evaluate RubricRanker on both deep research and RAG benchmarks and report the results in Tables~\ref{tab:main_results_dr} and~\ref{tab:main_results_rag}, respectively. We make the following observations:

\textbf{(1) RubricRanker demonstrates superior performance in both deep research and RAG scenarios.}
On deep research benchmarks, RubricRanker achieves the best average score of 60.1, outperforming the second-best baseline, Rank4Gen, by 2.6 points. On RAG benchmarks, RubricRanker also obtains the highest average EM score of 40.0, surpassing Rank4Gen by about 2 points. These consistent improvements demonstrate that rubric-based reranker training can select document sets that better support downstream LLM generation.

\textbf{(2) Generation-oriented rerankers do not generalize to deep research tasks.}
Generation-oriented rerankers clearly improve over vanilla rerankers on RAG benchmarks, where Rank4Gen outperforms BGE-Reranker-Large by 1.8 EM points on average. However, the gain nearly disappears on deep research benchmarks. This suggests that rerankers trained on RAG-style, closed-form tasks may not generalize well to open-ended deep research, highlighting the need for search-rubric-based training.

\textbf{(3) Reranking substantially improves generation quality over initial retrieval.}
All evaluated rerankers improve over initial retrieval in both scenarios: average scores increase from 54.0 to 56.1--60.1 on deep research benchmarks and from 34.3 to 35.3--40.0 on RAG benchmarks. This confirms that reranking retrieved documents generally benefits downstream generation.


\begin{table}[t]
\footnotesize
\centering
\setlength{\tabcolsep}{0.6mm}{
\begin{tabular}{lcccc}
\toprule
\textbf{Method} & \textbf{WebWalker.} & \textbf{Health.} & \textbf{HotpotQA} & \textbf{Avg.} \\ \midrule
RubricRanker & \textbf{58.0} & \textbf{61.5} & \textbf{38.0} & \textbf{52.5} \\ \midrule
\multicolumn{5}{l}{\textit{Training Stage}} \\
w/o RL & 55.0 & 61.0 & 37.2 & 51.1 \\
w/o SFT & 48.0 & 61.0 & 35.8 & 48.3 \\ \midrule
\multicolumn{5}{l}{\textit{SFT Label}} \\
w/o rubrics & 51.5 & 60.0 & 36.0 & 49.2 \\
Relevance Ranking & 50.0 & 59.2 & 35.0 & 48.1 \\ \bottomrule
\end{tabular}}
\caption{Ablation study of RubricRanker on WebWalkerQA, HealthBench, and HotpotQA. Avg. denotes the average score across the three datasets.}
\label{tab:ablation}
\end{table}

\subsection{Ablation Study} \label{sec:ablation}
We conduct ablation studies to evaluate the contributions of different components in two aspects: (1) the two-stage training framework and (2) label construction. Experiments are conducted on two deep research datasets, WebWalkerQA and HealthBench, and one RAG dataset, HotpotQA. The overall results are reported in Table~\ref{tab:ablation}. We summarize our findings as follows.

\textbf{(1) Training stage.}
We ablate the two-stage training framework by removing rubric-based RL (``w/o RL'') and cold-start SFT (``w/o SFT''). Removing RL decreases the average score from 52.5 to 51.1, showing that rubric-based reward optimization provides additional gains beyond SFT. Removing cold-start SFT leads to a larger performance drop, from 52.5 to 48.3, highlighting its critical role in providing a stable initialization for RL training.

\textbf{(2) Label construction.}
We further study how different SFT labels affect RubricRanker. Removing query-specific rubrics from the teacher's guidance during label construction (``w/o rubrics'') reduces the average score from 52.5 to 49.2, demonstrating that rubric guidance helps the teacher model identify document sets that better satisfy downstream generation requirements. We also test a widely used relevance-ranking label (``Relevance Ranking'')~\cite{rankvicuna,rankzephyr}, in which the teacher outputs a complete ranking ordered by document relevance rather than selecting a document set for generation. This variant further decreases the average score to 48.1, indicating that relevance ranking is insufficient for generation-oriented document-set selection and validating the necessity of rubric-guided label construction.

\begin{figure}[!t]
	\centering
	\includegraphics[width=1\linewidth]{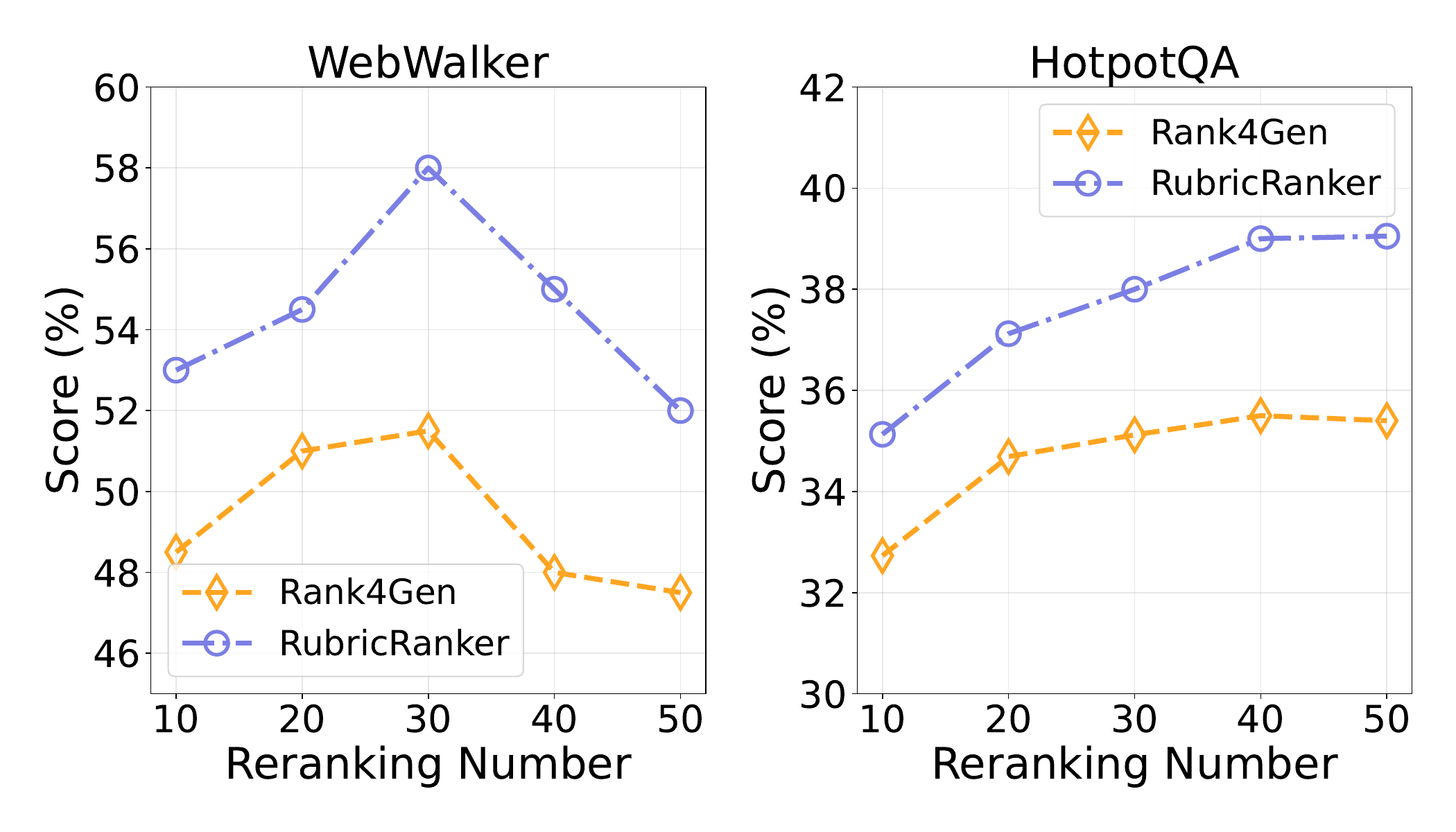}
	\caption{Performance of RubricRanker when reranking different numbers of retrieved documents on WebWalkerQA and HotpotQA.}
	\label{fig:rerank_num}
\end{figure}

\subsection{Different Numbers of Reranked Documents}

The number of retrieved documents provided to the reranker can affect downstream generation quality. In this section, we investigate how this factor influences final performance. Specifically, we evaluate RubricRanker on WebWalkerQA and HotpotQA while reranking the top 10, 20, 30, 40, and 50 retrieved documents. As shown in Figure~\ref{fig:rerank_num}, increasing the number of reranked documents generally improves downstream performance because relevant documents may appear at lower positions in the initial retrieval list. However, the gains saturate and may even decline beyond a certain threshold, such as 30 documents on WebWalkerQA and 40 documents on HotpotQA. This may occur because most relevant documents have already been retrieved, while reranking too many documents at once creates an overly long input context that can degrade the reranker's performance.

\begin{figure}[!t]
	\centering
	\includegraphics[width=0.99\linewidth]{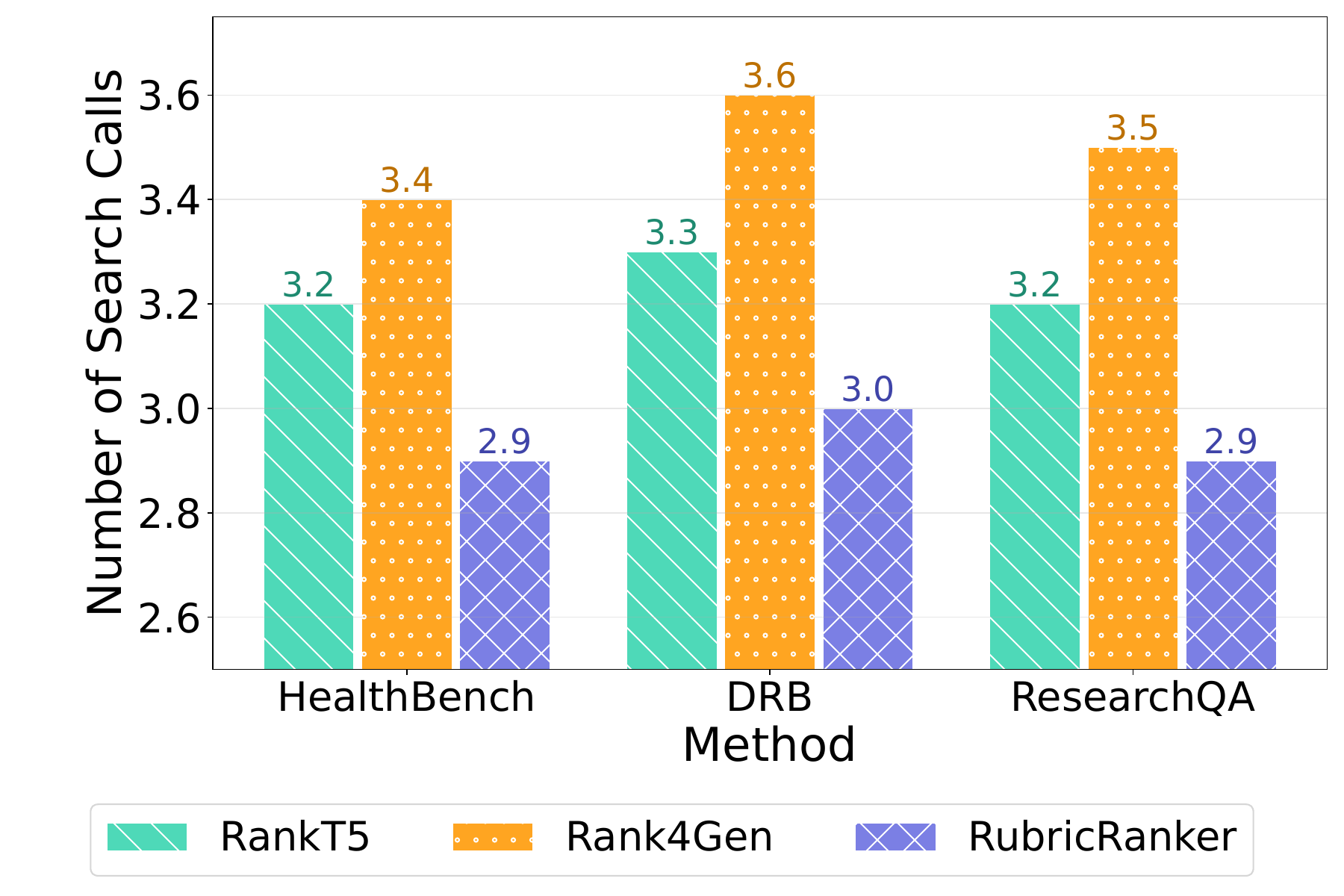}
	\caption{The number of search calls made by the Dr-Tulu deep research agent when using different document rerankers.}
	\label{fig:search_turns}
\end{figure}

\subsection{Search Call Analysis}
We further analyze the number of search calls made by the deep research agent when equipped with different document rerankers. Fewer search calls indicate that the reranker provides sufficient evidence more efficiently. We use Dr-Tulu (8B) as the deep research agent and evaluate it on HealthBench, DeepResearchBench (DRB), and ResearchQA. We compare RubricRanker with two competitive reranking baselines, RankT5 and Rank4Gen. As shown in Figure~\ref{fig:search_turns}, RubricRanker consistently leads to fewer search calls across all three datasets. On HealthBench, it reduces the average number of calls from 3.2 with RankT5 and 3.4 with Rank4Gen to 2.9, corresponding to relative reductions of 9.4\% and 14.7\%, respectively. On ResearchQA, it reduces search calls from 3.2 and 3.5 to 2.9, corresponding to reductions of 9.4\% and 17.1\%. These consistent improvements suggest that RubricRanker provides more useful evidence at each step, allowing the agent to satisfy its information needs earlier and avoid repeated searches.

\section{Conclusion}
We present RubricRanker, a document reranker that moves beyond individual query--document relevance matching by modeling query-specific requirements for high-quality document sets. Rather than supervising the model only with single-document relevance labels, we formulate hierarchical search rubrics that describe both the collective properties of a selected set and the quality of each document. Based on these rubrics, we propose a two-stage training framework that combines rubrics-guided SFT for cold-start training with rubric-based RL using fine-grained rubric rewards. The trained reranker internalizes these requirements and directly selects a high-quality subset from retrieved candidates at inference without rubric inputs. Experiments demonstrate consistent gains on deep research and RAG benchmarks, strong cross-scenario generalization, and fewer search calls for deep research agents.

\section*{Limitations} \label{limitation}
Despite the effectiveness of RubricRanker, our work still has several limitations. (1) The construction of rubric-based rewards relies on GPT-5.1, which can be costly when scaling to large training corpora. In the future, we plan to train a specialized reward model to provide more cost-effective and efficient rubric-based feedback. (2) Due to the high cost of LLM calls and the slow test-time inference of deep research agents, we sample a subset of queries for evaluation on deep research benchmarks. In future work, we will consider evaluating on the full benchmark sets to provide more comprehensive results. (3) Our evaluation still depends on the final generation quality of LLM agents. Even when RubricRanker selects a high-quality document set, the agent may still make mistakes during reasoning or generation. Developing more objective metrics to directly evaluate the quality of selected document sets is a promising direction.


\clearpage
\appendix

\section{Details of Deep Research Benchmarks} \label{sub:dr_datasets}
We provide brief descriptions of the deep research benchmarks used in our evaluation.

\paragraph{HealthBench.}
HealthBench~\cite{healthbench} is a healthcare-oriented benchmark that evaluates whether models can answer medically related user questions in a helpful, accurate, and safe manner. Its queries often require careful interpretation of user intent and reliable synthesis of evidence, making it suitable for evaluating long-form responses in high-stakes domains. We evaluate HealthBench with an adapted OpenAI simple-evals pipeline\footnote{\url{https://github.com/openai/simple-evals}}: each multi-turn case is flattened into a single doctor--patient conversation, and the model is instructed to answer based on this conversation. We use GPT-4.1 as the LLM judge. For efficiency, we randomly sample 100 cases for evaluation.

\paragraph{WebWalkerQA.}
WebWalkerQA~\cite{webwalkerqa} focuses on complex web and website-level question answering. Solving its questions requires agents to search, browse, and connect information distributed across web pages, rather than relying on a single retrieved passage. We evaluate answers with GPT-4.1 as the LLM judge under a unified evaluation pipeline, using a sampled test set of 200 queries.

\paragraph{DeepResearchBench.}
DeepResearchBench~\cite{deepresearchbench} evaluates general-domain deep research capabilities using open-ended questions. The benchmark emphasizes qualities such as comprehensiveness, depth of analysis, instruction following, and readability, which are assessed through rubric-based evaluation. We evaluate 50 English and 50 Chinese questions, score the generated articles on Comprehensiveness, Insight/Depth, Instruction-Following, and Readability, and report the macro average as the overall score. Following the official evaluation protocol~\cite{deepresearchbench}, we use Gemini 2.5 Flash as the judge and the Jina API to scrape URLs for evidence snippets when needed; for our system outputs, we use the URL contents collected by the corresponding search or browsing tools.

\paragraph{ResearchQA.}
ResearchQA~\cite{researchqa} contains research-style questions that require collecting external evidence and producing synthesized long-form answers. It tests whether an agent can identify useful information, organize evidence, and provide a coherent response to complex information needs. Following the original ResearchQA evaluation suite~\cite{researchqa}, we report average rubric scores on the 100-question subset used to evaluate deep research systems, with GPT-4.1-mini as the judge.

\section{Details of Baselines} \label{sub:baseline_details}
We provide additional details of the baseline rerankers used in our experiments.

\subsection{Vanilla Rerankers.}
These rerankers are primarily designed to estimate query--document relevance and improve the ranking of retrieved documents.

\paragraph{BGE-Reranker-Large.}\footnote{\url{https://huggingface.co/BAAI/bge-reranker-large}} BGE-Reranker-Large is a cross-encoder reranker from the BGE family. It jointly encodes a query and a candidate document to produce a relevance score, and is widely used as a strong open-source reranking baseline.

\paragraph{MonoT5}~\cite{monot5}. MonoT5 formulates passage reranking as a sequence-to-sequence task with T5, where the model predicts relevance-oriented outputs for each query--document pair and uses them to score candidates.

\paragraph{RankT5}~\cite{rankt5}. RankT5 also builds on the T5 architecture, but is optimized with ranking-oriented objectives to produce effective pointwise relevance estimates for document reranking.

\paragraph{RankVicuna}~\cite{rankvicuna}. RankVicuna is an instruction-tuned listwise reranker based on Vicuna. It is trained with ranked lists generated by ChatGPT, enabling it to reorder candidate documents through list-level comparison.

\paragraph{RankZephyr}~\cite{rankzephyr}. RankZephyr is a listwise reranker distilled from GPT-4. It adopts a two-stage training strategy and applies a sliding-window strategy to handle longer candidate lists.

\subsection{Generation-oriented Rerankers.}
These rerankers are designed to optimize downstream LLM answer generation, where the goal is not only to rank individually relevant documents but also to select a high-quality document set that improves generated answers.

\paragraph{SetR}~\cite{setr}. SetR shifts document ranking toward set-wise passage selection for RAG. It identifies the information requirements of a query through reasoning and selects a passage set that collectively satisfies these requirements.

\paragraph{Rank4Gen}~\cite{rank4gen}. Rank4Gen is a generator-aware reranker that aligns document selection with downstream generation quality. Instead of optimizing only relevance, it trains the ranker with signals derived from response quality and models generator-specific preferences.

\section{Implementation Details} \label{app:implementation_details}
We provide additional implementation details of RubricRanker to facilitate reproducibility. The code is available at \url{https://github.com/8421BCD/RubricRanker}.

\subsection{Training Query Construction}
We construct training queries from both deep research and RAG scenarios. For deep research datasets, we run Dr-Tulu-8B~\cite{Dr-tulu} to generate full trajectories for each original question and randomly sample two agent queries from each trajectory. For RAG datasets, we randomly sample user questions from the original training sets. Table~\ref{tab:training_queries} reports the number of training queries used in the SFT and RL stages.

\begin{table}[t]
\footnotesize
\centering
\setlength{\tabcolsep}{2mm}{
\begin{tabular}{lccc}
\toprule
\textbf{Dataset} & \textbf{SFT} & \textbf{RL} & \textbf{All} \\ \midrule
\multicolumn{4}{l}{\textit{\textbf{Deep Research}}} \\
OpenScholar & 2,562 & 3,671 & 6,233 \\
SearchArena & 1,628 & 2,172 & 3,800 \\
GlaiveAI-Reasoning-v1-20M & 880 & 1,265 & 2,145 \\
WebWalker-Silver & 884 & 1,516 & 2,400 \\ \midrule
\multicolumn{4}{l}{\textit{\textbf{RAG}}} \\
HotpotQA & 1,941 & 3,000 & 4,941 \\
NQ & 1,948 & 3,000 & 4,948 \\ \midrule
All & 9,843 & 14,624 & 24,467 \\ \bottomrule
\end{tabular}}
\caption{Number of training queries used for SFT and RL. Deep research datasets use sampled agent queries, while RAG datasets use sampled user questions.}
\label{tab:training_queries}
\end{table}

\subsection{Cold-Start SFT Stage}
\begin{figure*}[!tb]
  \centering
  \includegraphics[width=1\linewidth]{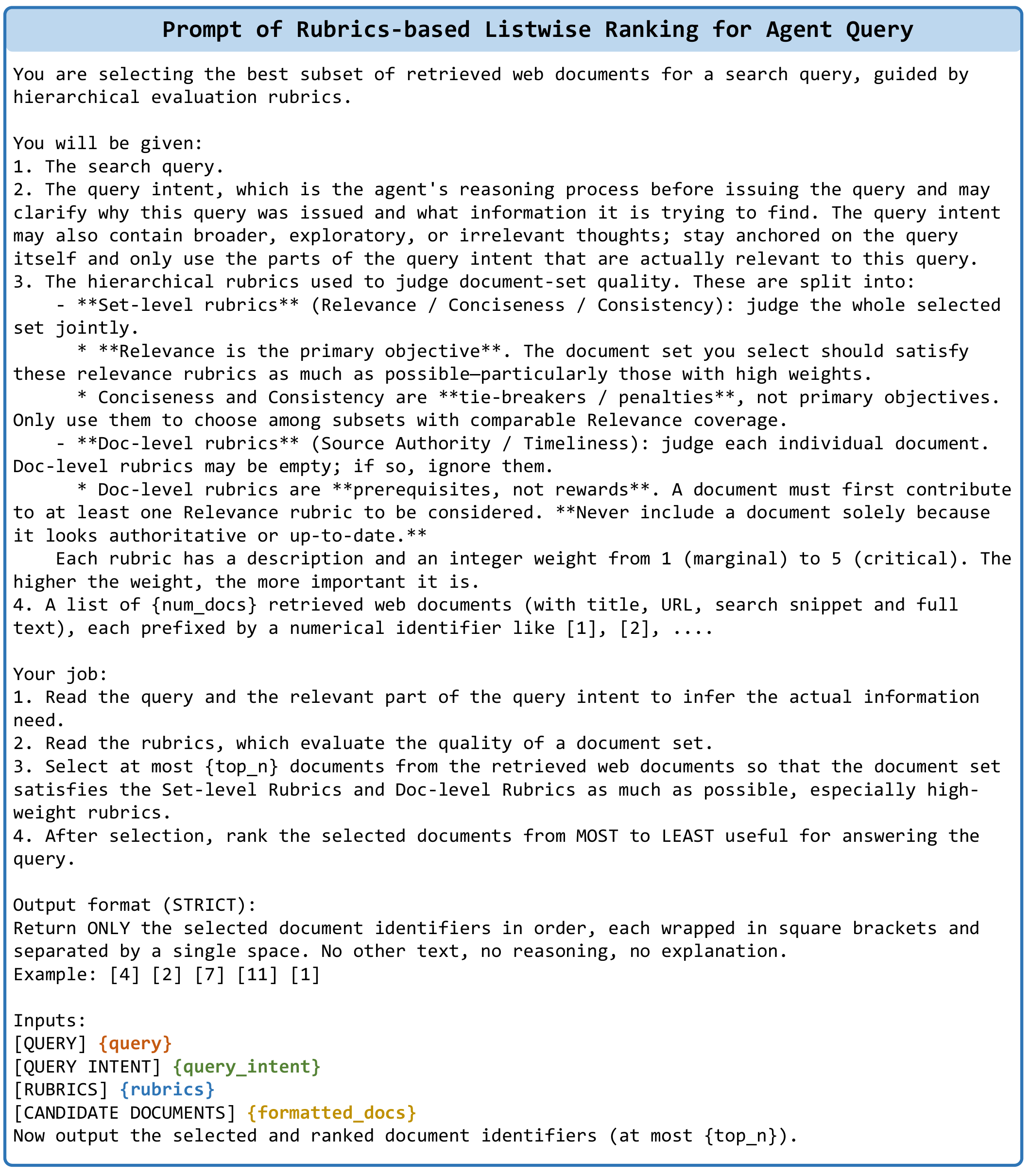}
  \caption{The prompt for constructing silver labels for an agent sub-query in the deep research scenario. The input consists of the query, query intent (agent reasoning), candidate documents, and query-specific rubrics.}
  \label{fig:rubrics_based_listwise_rerank_subq}
\end{figure*}

\begin{figure*}[!tb]
  \centering
  \includegraphics[width=1\linewidth]{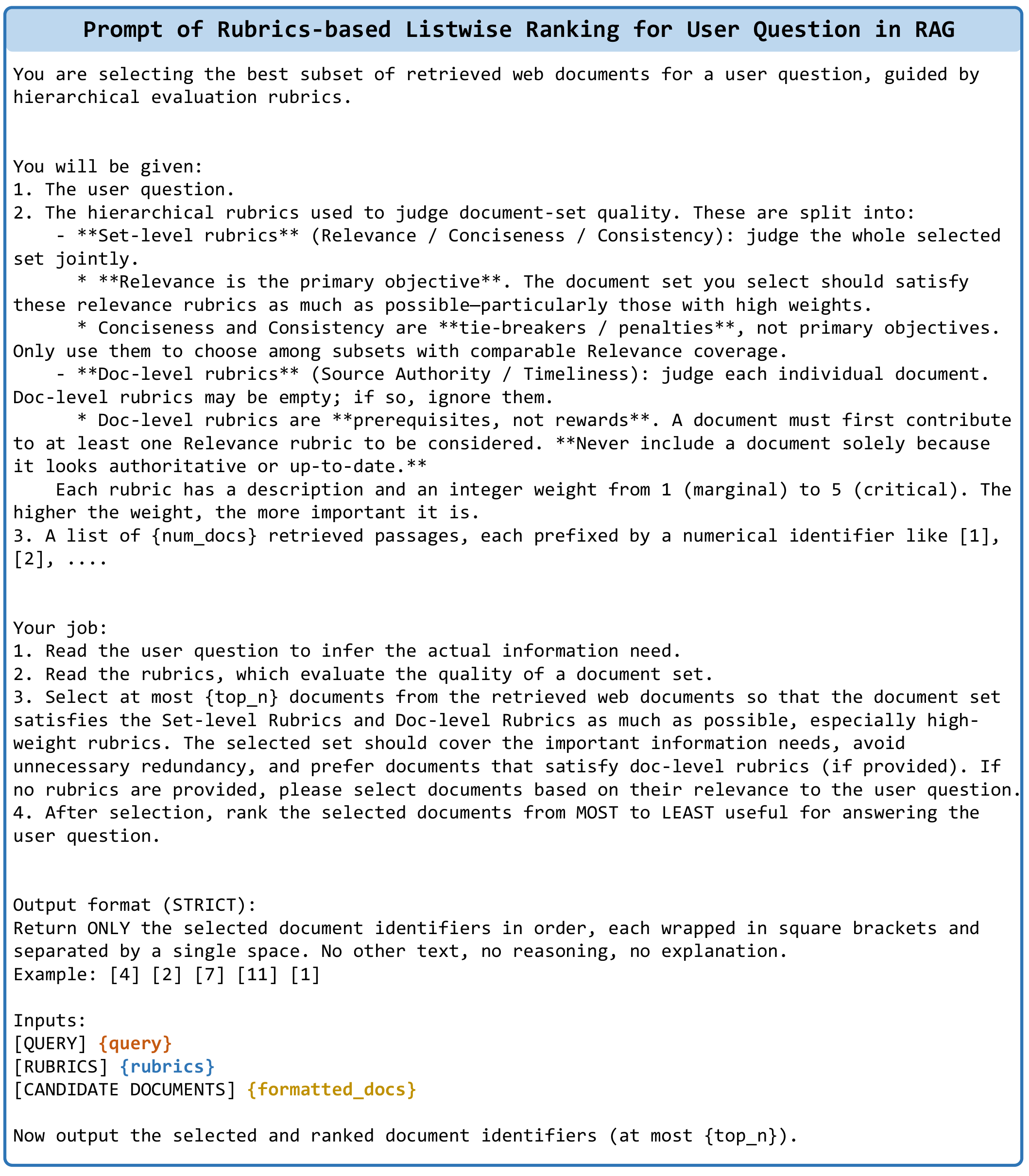}
  \caption{The prompt for constructing silver labels for a user question in the RAG scenario. The input consists of the user question, candidate documents, and query-specific rubrics.}
  \label{fig:rubrics_based_listwise_rerank_userq}
\end{figure*}

In the cold-start SFT stage, we first use GPT-5.1 to construct silver labels for rubric-based listwise reranking. The prompts for agent queries in the deep research scenario and user questions in the RAG scenario are shown in Figures~\ref{fig:rubrics_based_listwise_rerank_subq} and~\ref{fig:rubrics_based_listwise_rerank_userq}, respectively. We then train the backbone LLM, Qwen3-8B,\footnote{\url{https://huggingface.co/Qwen/Qwen3-8B}} with LlamaFactory~\cite{llamafactory}. The model input follows the same scenario-specific format, consisting of the query and candidate documents, with additional query intent for agent queries. We set the learning rate to 5e-6, the per-GPU batch size to 1, and the number of gradient accumulation steps to 8. Training is accelerated with DeepSpeed ZeRO-2~\cite{deepspeed} and FlashAttention-2~\cite{flashattention}. We use BF16 mixed precision and train the model for 2 epochs.

\subsection{Rubric-Based RL Stage} \label{subsec:rl}
After cold-start SFT, we further optimize RubricRanker with the GRPO reinforcement learning algorithm~\cite{grpo} implemented in the VERL framework\footnote{\url{https://github.com/volcengine/verl}}.

During GRPO training, for each input $x$, we sample a group of output sequences $G = \{y_1, y_2, \ldots, y_G\}$. Each sequence $y_i$ is assigned a reward $r_i$, and rewards are normalized within the group $G$ to obtain the advantages $\hat{A}_i$. The token-level objective is defined as:
\begin{equation} \label{eq:grpo}
\begin{aligned}
\mathcal{J}_{\text{GRPO}}(\theta) &= \frac{1}{|G|} \sum_{i=1}^{|G|} \frac{1}{|y_i|} \sum_{t=1}^{|y_i|} \min \left(r_{i,t}(\theta) \hat{A}_{i,t}, \right. \\
&\quad \left. \text{clip} \left( r_{i,t}(\theta), 1 - \epsilon, 1 + \epsilon \right) \hat{A}_{i,t} \right) - \beta D_{\text{KL}}, \\
r_{i,t}(\theta) &= \frac{\pi_\theta(y_{i,t} \mid x, y_{i,<t})}{\pi_{\text{old}}(y_{i,t} \mid x, y_{i,<t})}, \\
D_{\text{KL}} &= D_{\text{KL}}(\pi_\theta \parallel \pi_{\text{ref}}),
\end{aligned}
\end{equation}
where $\epsilon$ and $\beta$ are hyperparameters.

\begin{figure*}[!tb]
  \centering
  \includegraphics[width=1\linewidth]{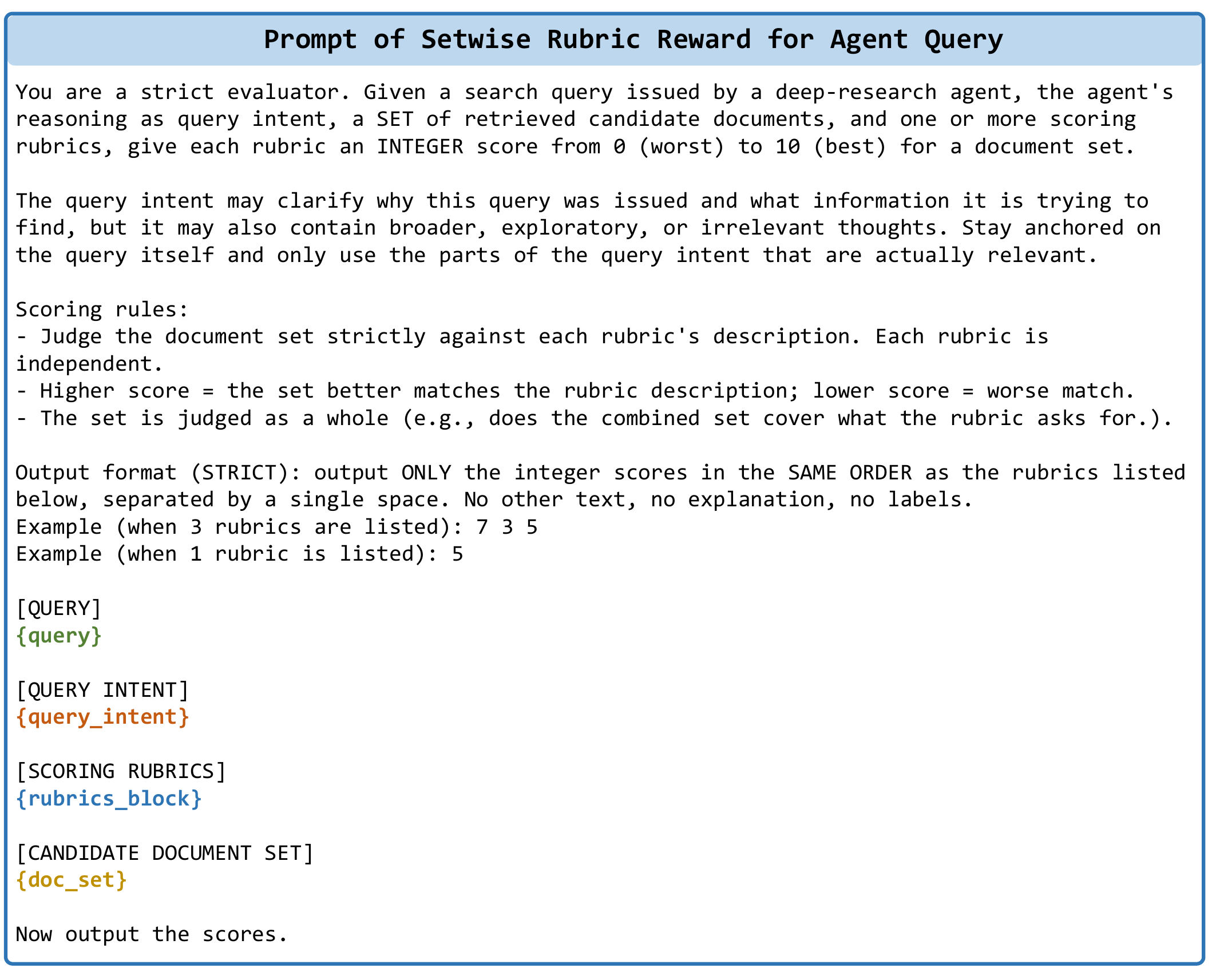}
  \caption{The prompt for set-level reward scoring for an agent sub-query in the deep research scenario. The input consists of the query, query intent (agent reasoning), the selected document set, and all set-level rubrics.}
  \label{fig:set_level_reward_subq}
\end{figure*}

\begin{figure*}[!tb]
  \centering
  \includegraphics[width=1\linewidth]{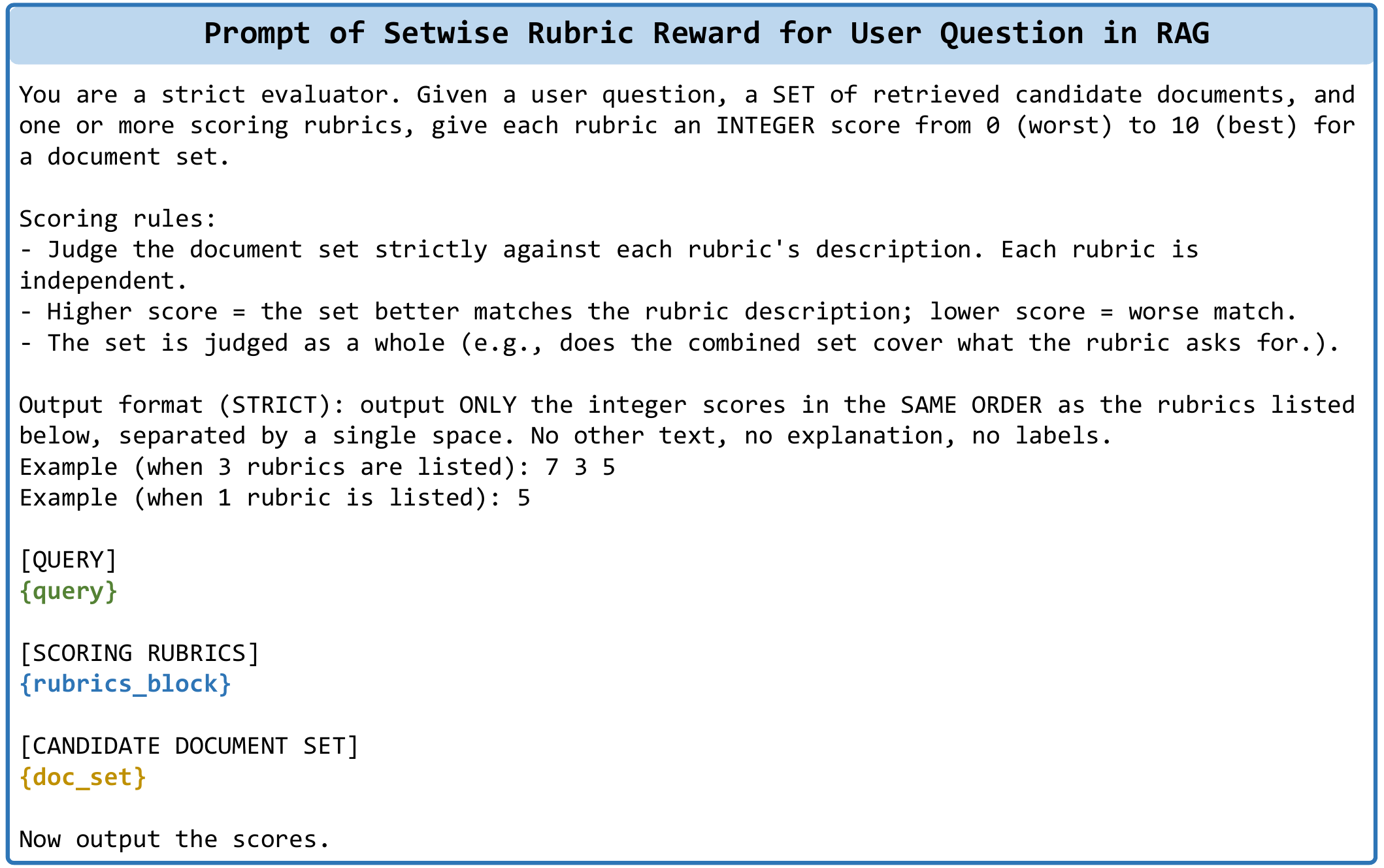}
  \caption{The prompt for set-level reward scoring for a user question in the RAG scenario. The input consists of the user question, the selected document set, and all set-level rubrics.}
  \label{fig:set_level_reward_userq}
\end{figure*}

\begin{figure*}[!tb]
  \centering
  \includegraphics[width=1\linewidth]{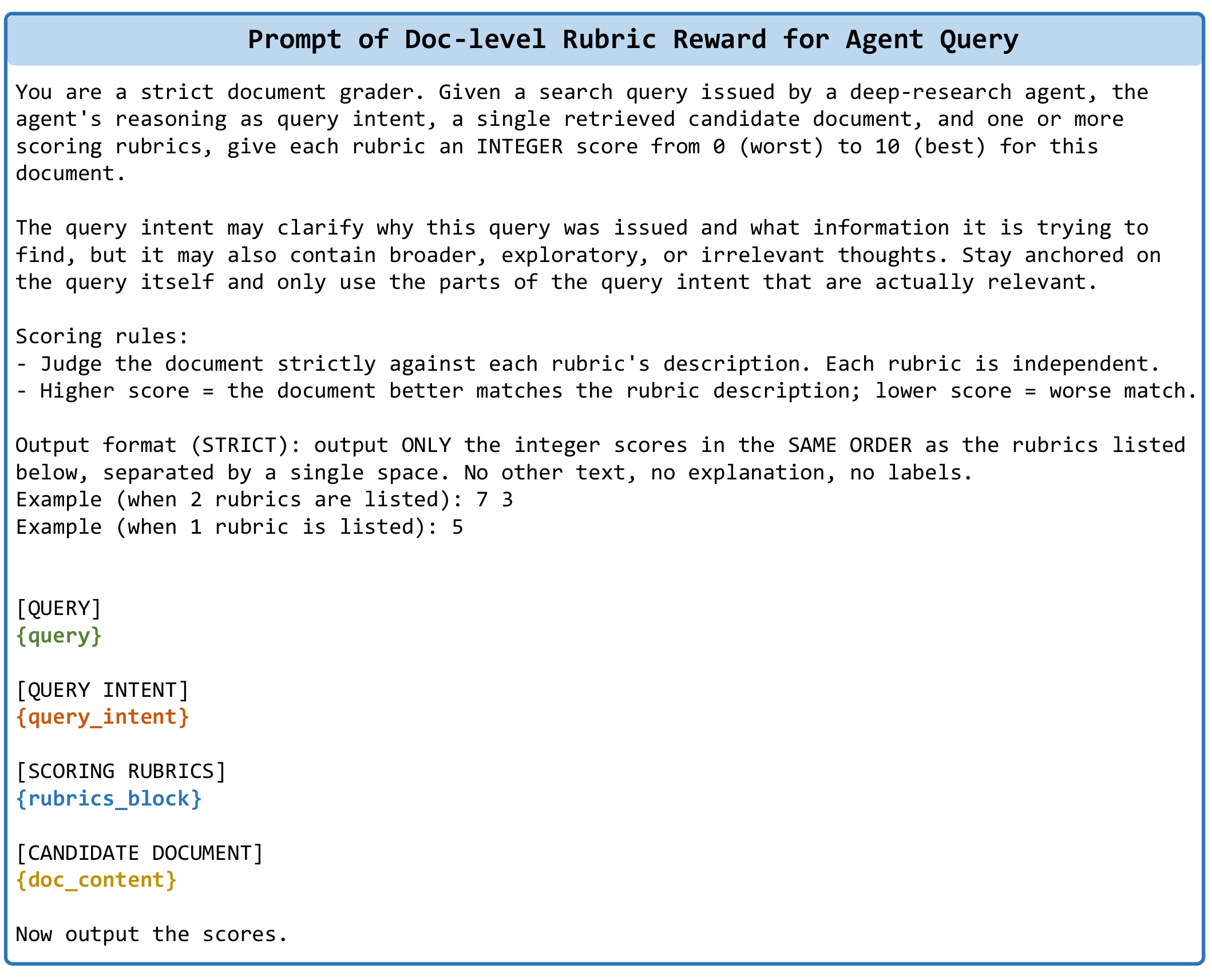}
  \caption{The prompt for document-level reward scoring for an agent sub-query in the deep research scenario. The input consists of the query, query intent (agent reasoning), a candidate document, and all document-level rubrics.}
  \label{fig:doc_level_reward_subq}
\end{figure*}

\begin{figure*}[!tb]
  \centering
  \includegraphics[width=1\linewidth]{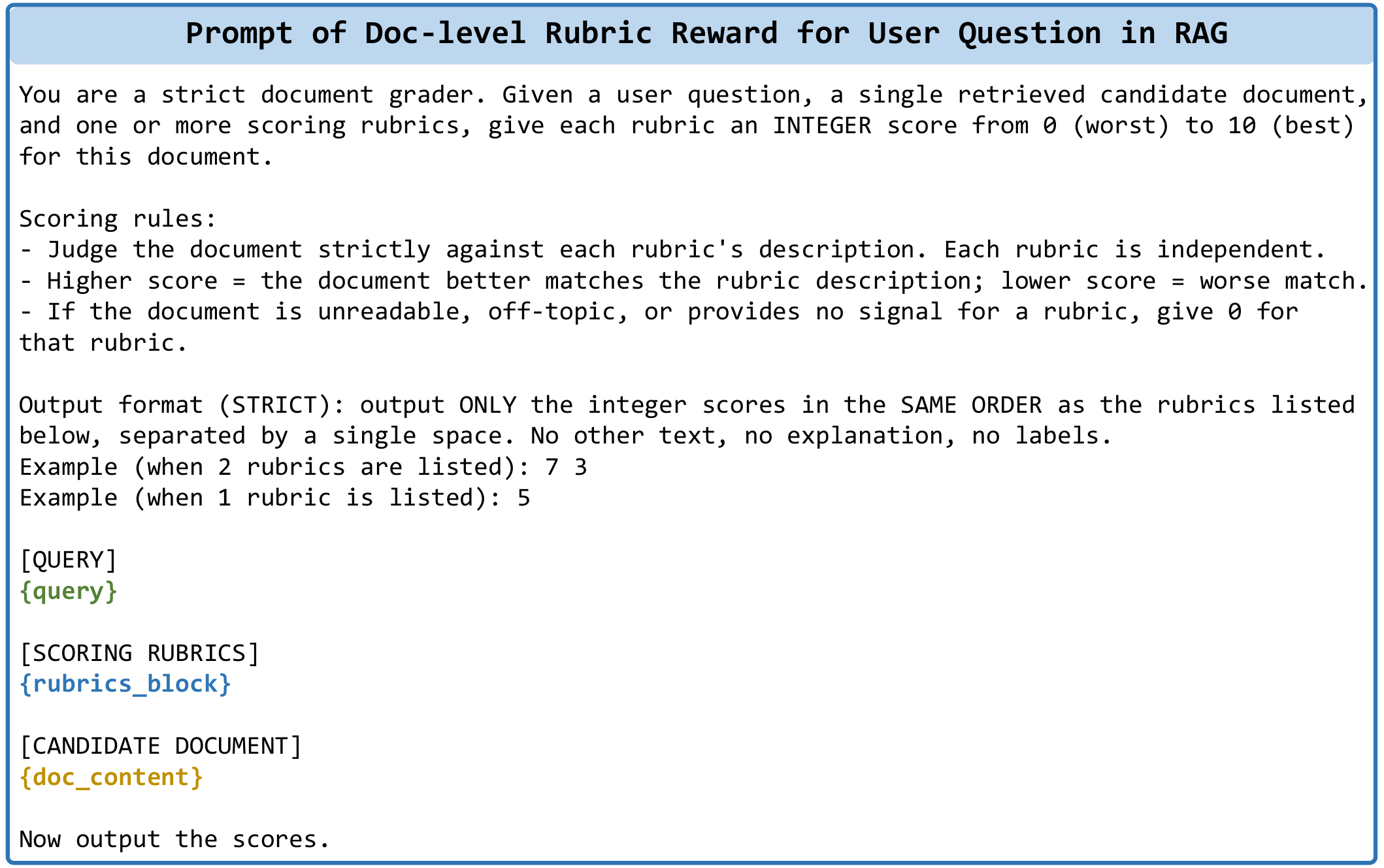}
  \caption{The prompt for document-level reward scoring for a user question in the RAG scenario. The input consists of the user question, a candidate document, and all document-level rubrics.}
  \label{fig:doc_level_reward_userq}
\end{figure*}

To compute set-level rubric rewards efficiently and reduce API cost, we feed all set-level rubrics for each rollout to the LLM judge in a single call and ask it to return the score for each rubric. The prompts for agent queries in the deep research scenario and user questions in the RAG scenario are shown in Figure~\ref{fig:set_level_reward_subq} and Figure~\ref{fig:set_level_reward_userq}, respectively. For document-level rubric rewards, we similarly score all document-level rubrics for each document in one judge call. The corresponding prompts for deep research agent queries and RAG user questions are shown in Figure~\ref{fig:doc_level_reward_subq} and Figure~\ref{fig:doc_level_reward_userq}.

The training batch size is set to 16, the mini-batch size is set to 8, and the number of rollouts per sample is 8. The model is trained for 150 steps on 8 NVIDIA H20 GPUs, with the learning rate set to 1e-6.

\section{Case Study}
We present two case studies to illustrate the query-specific rubrics used by RubricRanker. Table~\ref{tab:case_study_dr} shows an agent query from the deep research scenario, while Table~\ref{tab:case_study_rag} shows a user question from the RAG scenario.

\begin{table*}[t]
\scriptsize
\centering
\setlength{\tabcolsep}{1.4mm}
\renewcommand{\arraystretch}{1.16}
\begin{tabular}{p{0.10\linewidth}p{0.11\linewidth}p{0.72\linewidth}c}
\toprule
\multicolumn{4}{p{0.98\linewidth}}{\textbf{Query:} motion dazzle human perception} \\ \midrule
\textbf{Level} & \textbf{Category} & \textbf{Description} & \textbf{Weight} \\ \midrule
Set-level & Relevance & Documents clearly define motion dazzle and distinguish it from other forms of camouflage (e.g., crypsis, background matching), including its historical use in WWI ship dazzle camouflage and its proposed function in biology as disrupting motion judgments rather than hiding objects. & 4 \\
Set-level & Relevance & Documents provide empirical or review evidence on how motion dazzle affects human speed perception, including that dazzle patterns can both increase and decrease perceived speed depending on internal pattern motion, contrast, and spatial frequency, rather than claiming a single uniform effect. & 5 \\
Set-level & Relevance & Documents address how motion dazzle impacts perceived direction and trajectory (e.g., biases toward stripe orientation, altered interception/capture performance), linking oriented patterns on moving targets to systematic misjudgments of movement direction or path. & 4 \\
Set-level & Relevance & Documents explain mechanistic accounts from human motion vision that can generate motion dazzle effects, such as the aperture problem, ambiguity of local oriented edge motion, conflict between global object motion and internal texture motion, and dependence on contrast and spatial frequency. & 5 \\
Set-level & Relevance & Documents explicitly connect motion dazzle phenomena to broader principles or heuristics of human motion perception (e.g., pooling of local motion signals, non-veridical reconstruction of 2D motion, interactions between form and motion processing), rather than only describing behavioral effects. & 3 \\
Set-level & Conciseness & The document set minimizes redundancy by avoiding multiple papers or pages that restate the same core experimental findings or definitions of motion dazzle, while still covering definition, speed effects, direction/trajectory effects, and mechanistic explanations without including tangential material (e.g., general camouflage history with no motion-perception content). & 3 \\
Set-level & Consistency & Across the document set, key claims about motion dazzle (its definition, that it can both increase and decrease perceived speed depending on pattern parameters, and that it disrupts motion judgments rather than simple detection) are mutually compatible or, where differing results exist, the differences are clearly attributed to experimental conditions rather than left as unresolved contradictions. & 4 \\
Document-level & Source Authority & The document is directly relevant to motion dazzle and human motion perception and is published in or hosted by an authoritative scientific source such as a peer-reviewed vision science/biology/psychology journal, an academic publisher, or a reputable scientific repository (e.g., PubMed/PMC, major university site), providing empirical data or well-cited reviews rather than unsourced popular summaries. & 4 \\ \bottomrule
\end{tabular}
\caption{Case study of query-specific rubrics for an agent query in the deep research scenario. The original question is ``To what extent and how can the motion dazzle effect give significant insight into the general algorithms or heuristics that underlie general perception in humans?'' Empty document-level rubrics are omitted.}
\label{tab:case_study_dr}
\end{table*}

\begin{table*}[t]
\scriptsize
\centering
\setlength{\tabcolsep}{1.4mm}
\renewcommand{\arraystretch}{1.16}
\begin{tabular}{p{0.10\linewidth}p{0.11\linewidth}p{0.72\linewidth}c}
\toprule
\multicolumn{4}{p{0.98\linewidth}}{\textbf{Query:} What are the names of the current members of the American heavy metal band that wrote the music for Hurt Locker The Musical?} \\ \midrule
\textbf{Level} & \textbf{Category} & \textbf{Description} & \textbf{Weight} \\ \midrule
Set-level & Relevance & At least one document explicitly identifies the American heavy metal band Metallica as the group that wrote the music for ``Hurt Locker The Musical'' and clearly links the musical to Metallica. & 5 \\
Set-level & Relevance & The document set lists the current members of Metallica by name and role, including James Hetfield, Lars Ulrich, Kirk Hammett, and Robert Trujillo, making clear that these are the present lineup members. & 5 \\
Set-level & Conciseness & The document set contains mostly documents that either (a) identify Metallica as the band associated with ``Hurt Locker The Musical'' or (b) list Metallica's current members, with minimal repetition of the same member list and little content about unrelated bands, albums, or other musicals. & 3 \\
Set-level & Consistency & Across the document set, the identified band and its current members are consistent, without conflicting claims about which band wrote the music for ``Hurt Locker The Musical'' or disagreements over the current Metallica lineup (e.g., all agree on Hetfield, Ulrich, Hammett, and Trujillo and do not present alternative current members). & 4 \\ \bottomrule
\end{tabular}
\caption{Case study of query-specific rubrics for a user question in the RAG scenario. Empty document-level rubrics are omitted.}
\label{tab:case_study_rag}
\end{table*}

The deep research case demonstrates that our rubrics capture multi-aspect evidence needs beyond topical relevance. The generated criteria require documents to cover definitions, empirical findings, perceptual mechanisms, broader theoretical implications, and source authority, all of which are important for open-ended scientific questions. In contrast, the RAG case focuses on a smaller set of highly targeted requirements: identifying the band, listing the current members, avoiding redundant or off-topic evidence, and ensuring consistency. These examples show that query-specific rubrics adapt to different information needs and provide fine-grained guidance for selecting document sets.

\section{Use of AI Assistants}
We use ChatGPT to improve the presentation of this paper.\footnote{\url{https://chatgpt.com/}}

\begin{figure*}[!tb]
  \centering
  \includegraphics[width=1\linewidth]{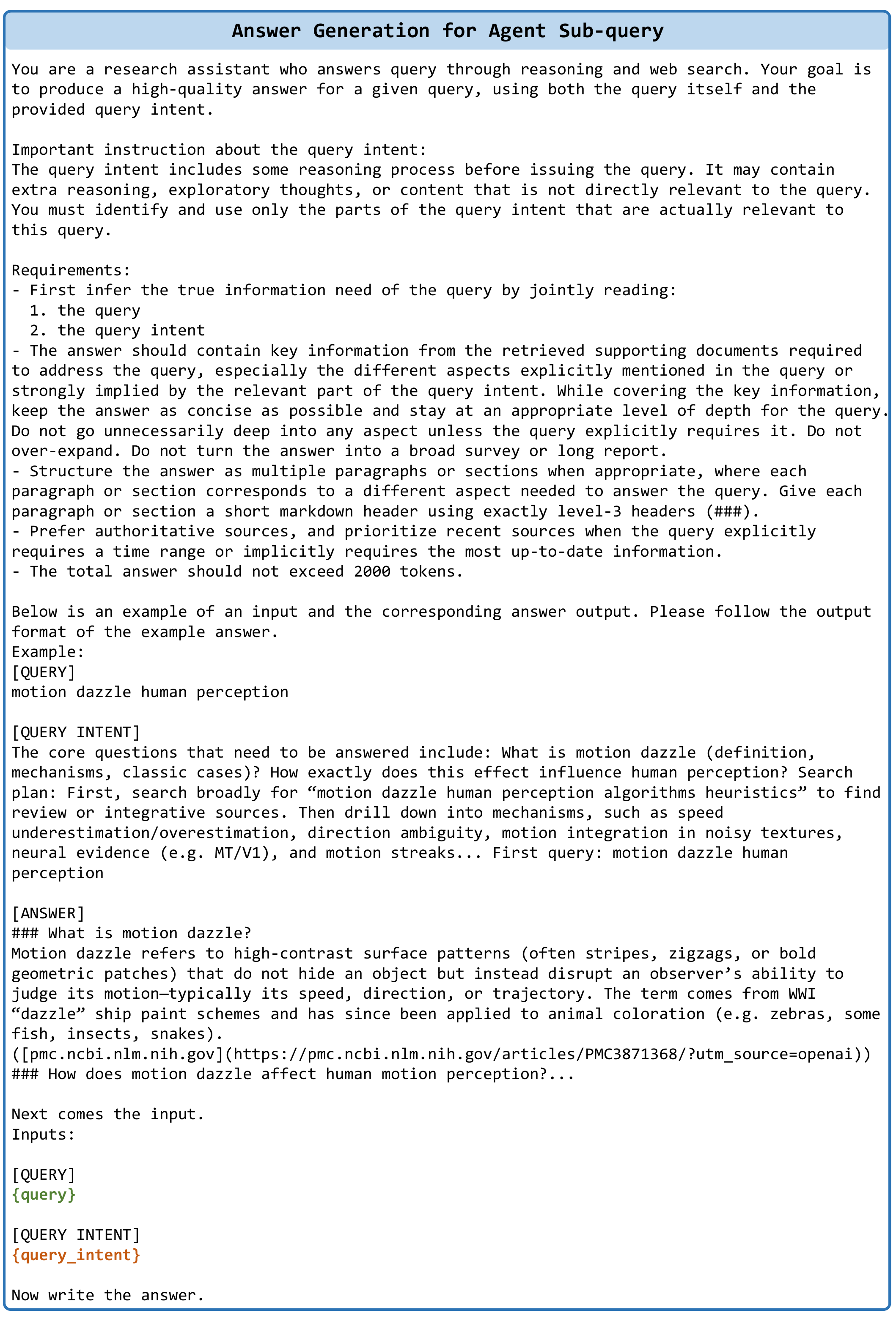}
  \caption{The prompt for generating an answer to an agent sub-query. The input consists of the query and query intent (agent reasoning).}
  \label{fig:answer_generation}
\end{figure*}


\begin{figure*}[!tb]
  \centering
  \includegraphics[width=1\linewidth]{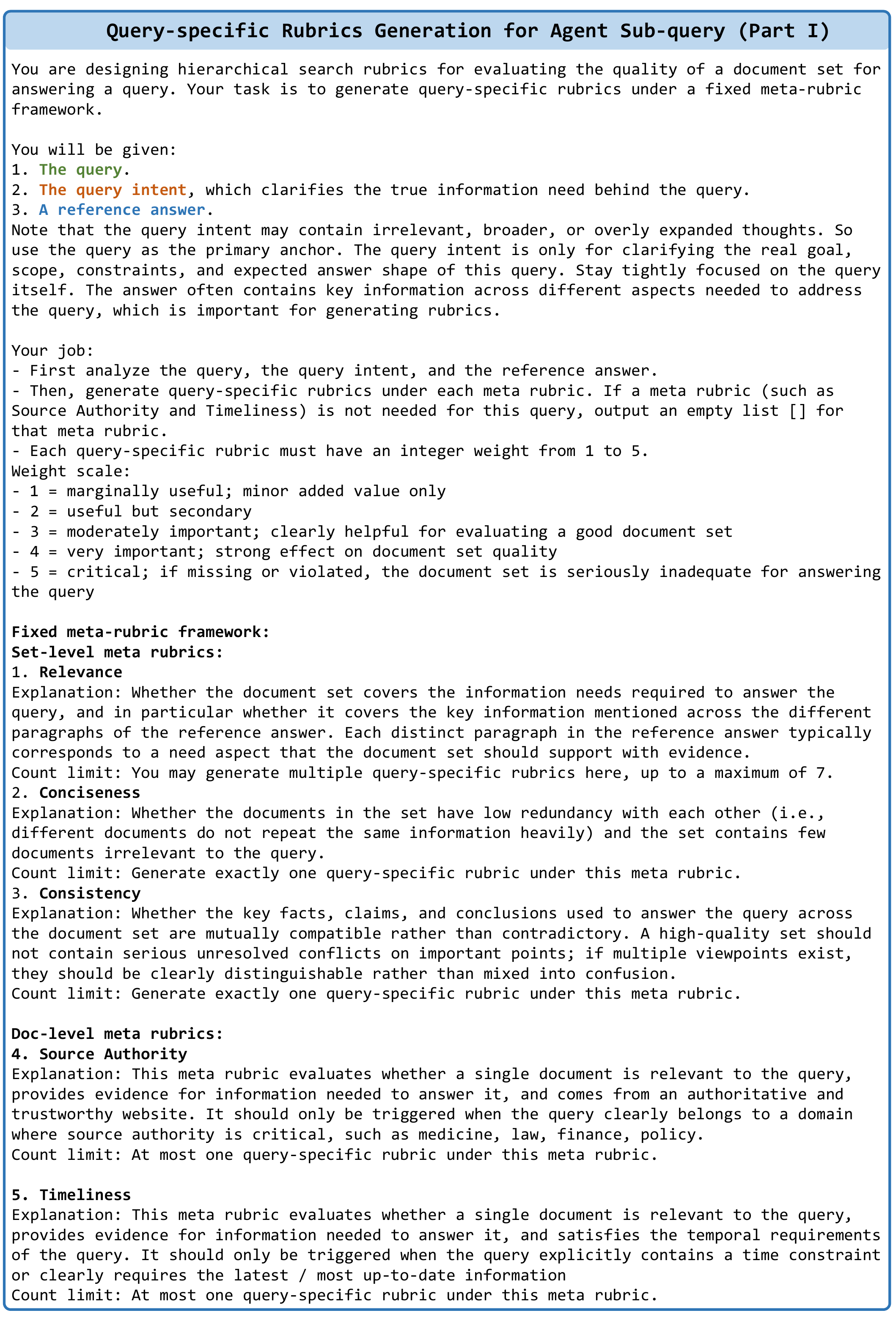}
  \caption{The prompt for generating query-specific rubrics for an agent sub-query (Part I). The input consists of the query, query intent (agent reasoning), and reference answer.}
  \label{fig:rubrics_generation_subq-1}
\end{figure*}

\begin{figure*}[!tb]
  \centering
  \includegraphics[width=1\linewidth]{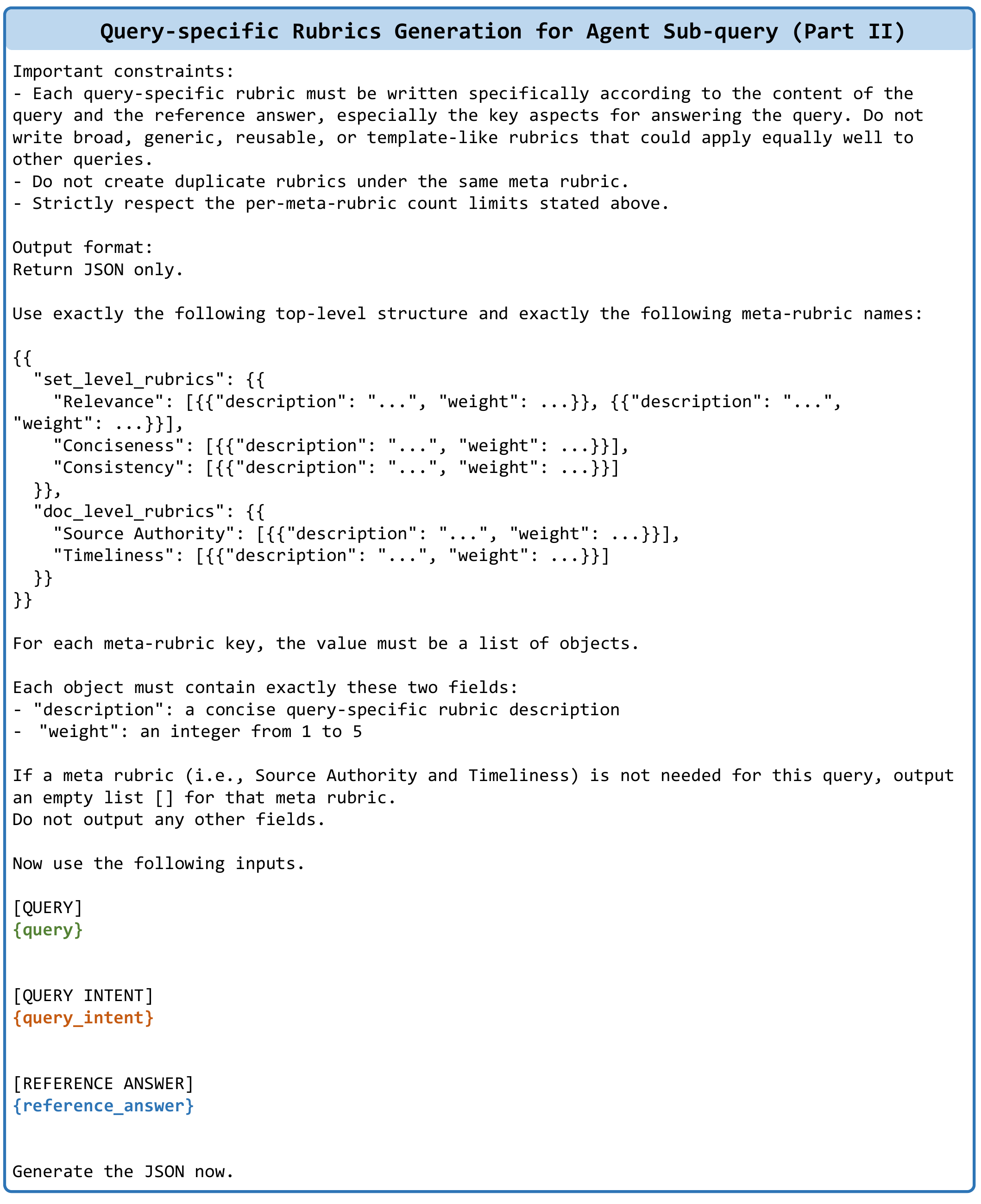}
  \caption{The prompt for generating query-specific rubrics for an agent sub-query (Part II). The input consists of the query, query intent (agent reasoning), and reference answer.}
  \label{fig:rubrics_generation_subq-2}
\end{figure*}


\begin{figure*}[!tb]
  \centering
  \includegraphics[width=1\linewidth]{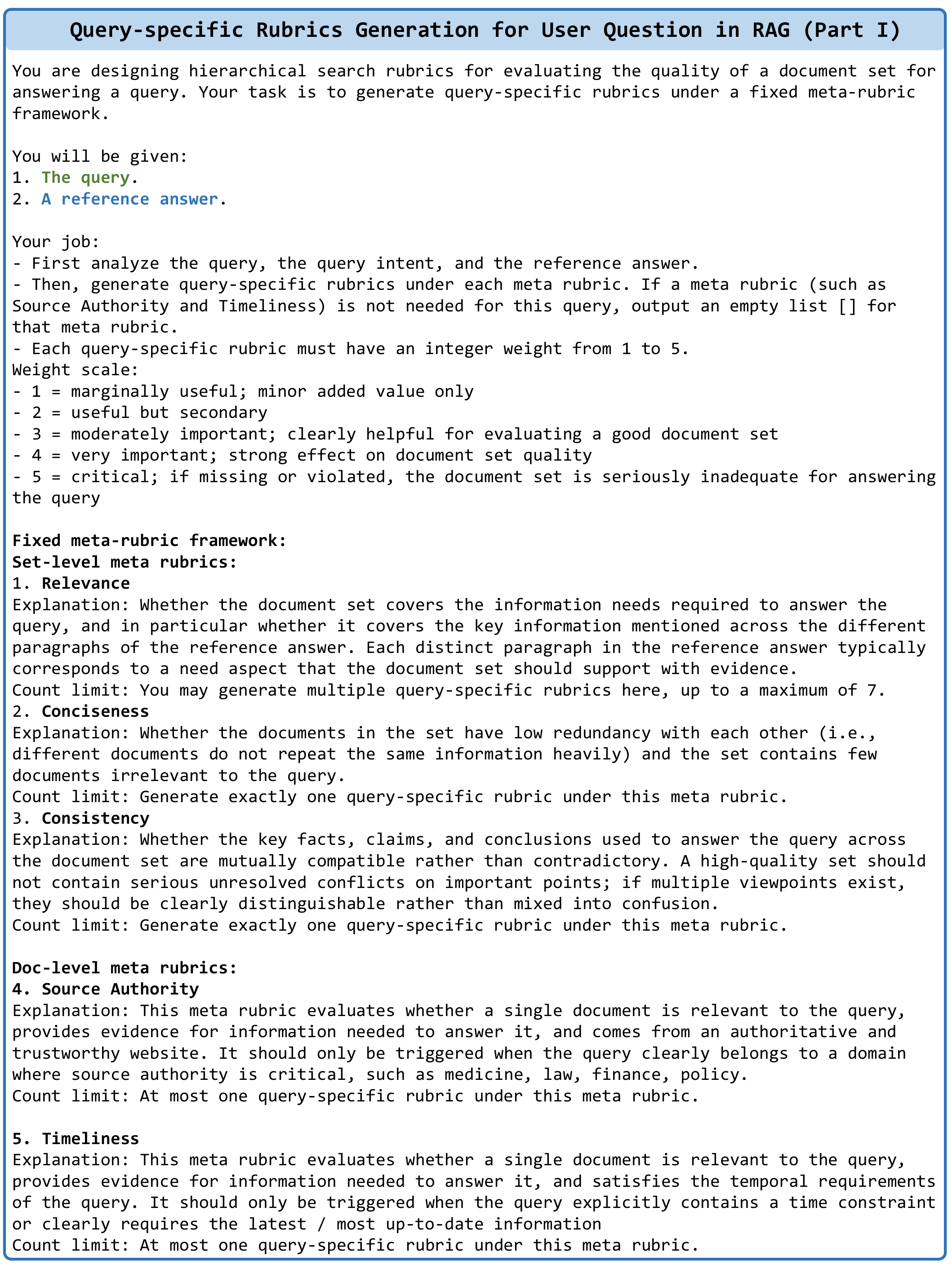}
  \caption{The prompt for generating query-specific rubrics for a user question in the RAG scenario (Part I). The input consists of the query and gold answers.}
  \label{fig:rubrics_generation_userq-1}
\end{figure*}

\begin{figure*}[!tb]
  \centering
  \includegraphics[width=1\linewidth]{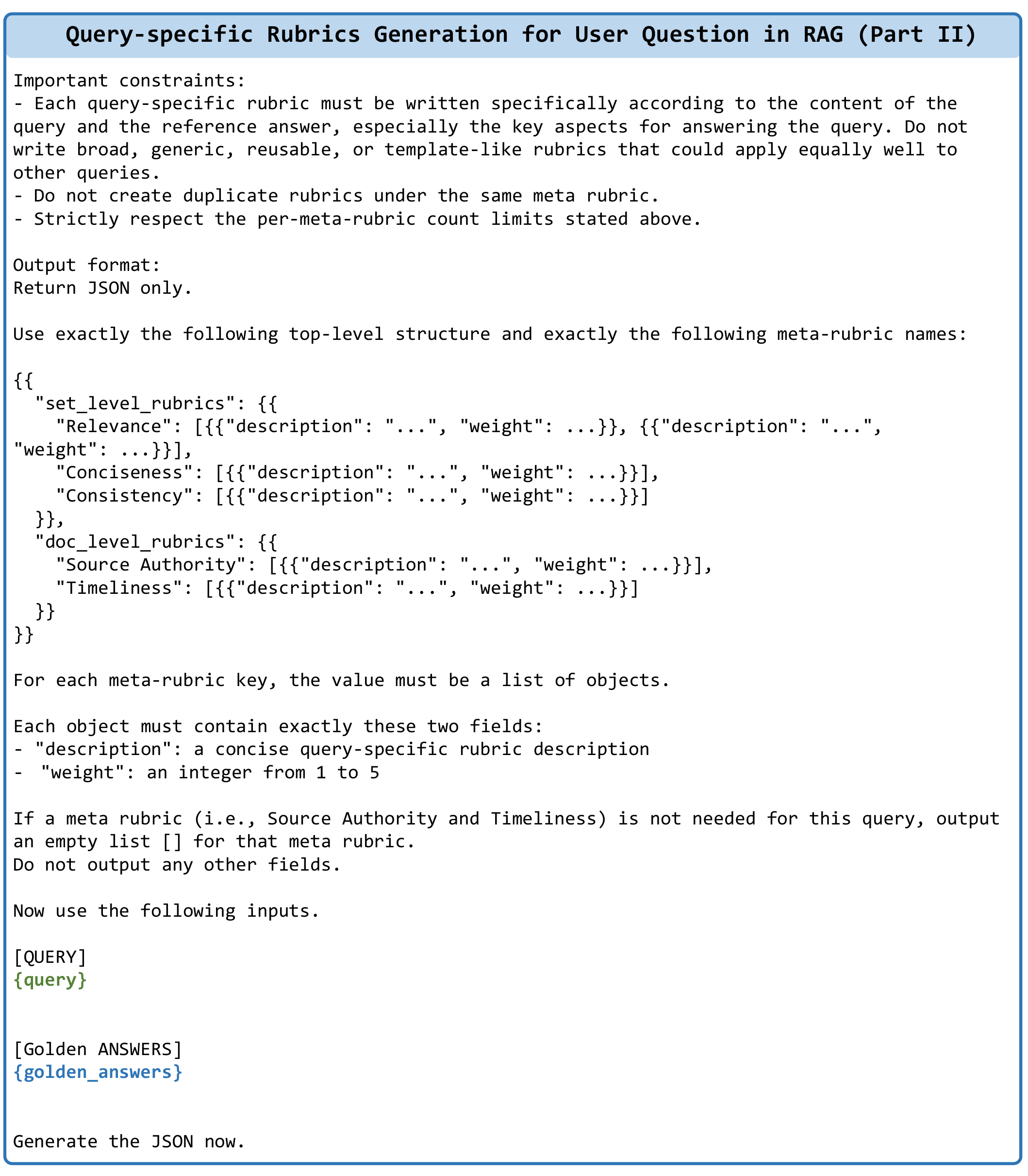}
  \caption{The prompt for generating query-specific rubrics for a user question in the RAG scenario (Part II). The input consists of the query and gold answers.}
  \label{fig:rubrics_generation_userq-2}
\end{figure*}

\begin{figure*}[!tb]
  \centering
  \includegraphics[width=1\linewidth]{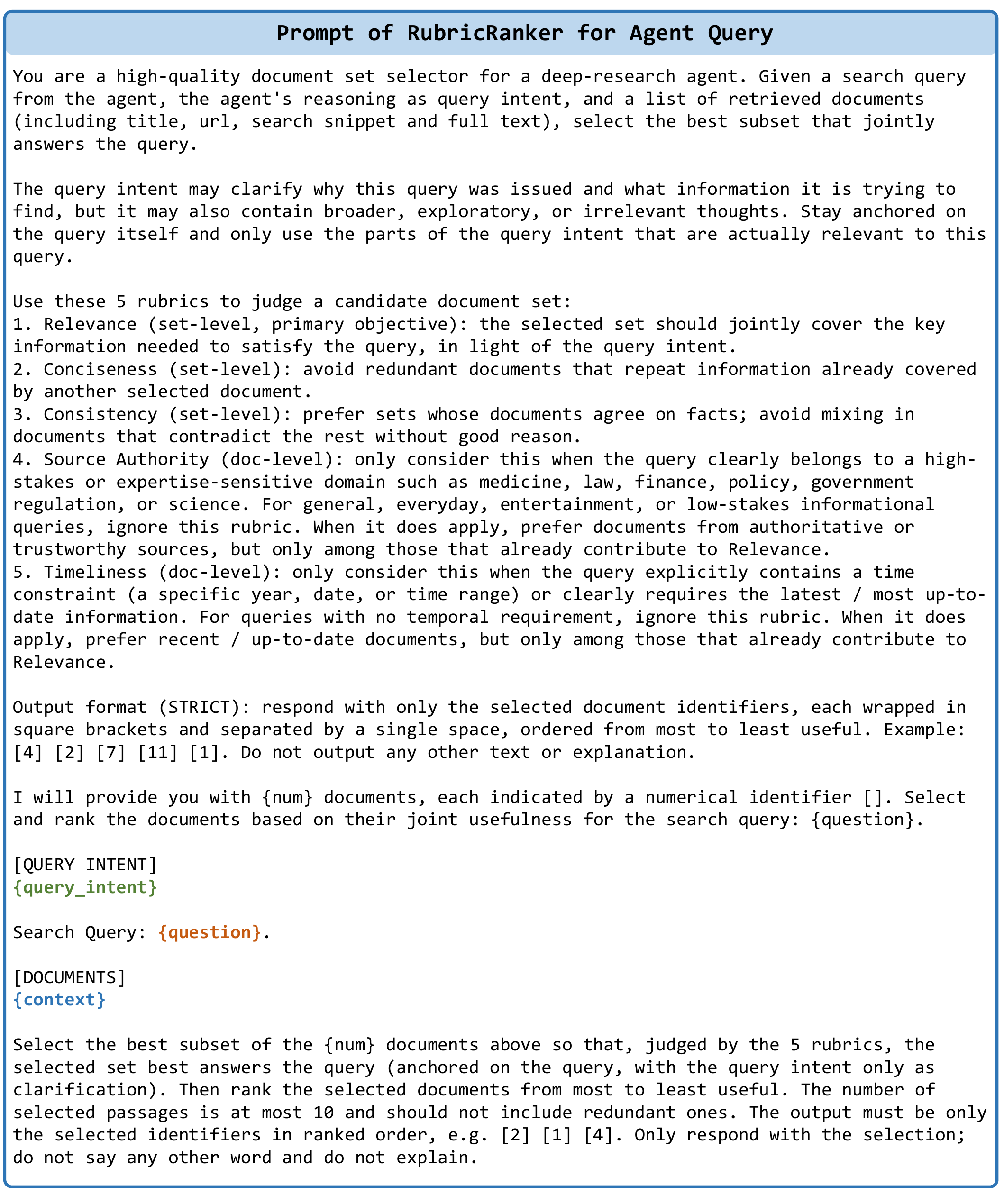}
  \caption{The RubricRanker prompt for an agent sub-query in the deep research scenario. The input consists of the agent query, query intent (agent reasoning), and candidate documents.}
  \label{fig:rubricrank_prompt_subq}
\end{figure*}

\begin{figure*}[!tb]
  \centering
  \includegraphics[width=1\linewidth]{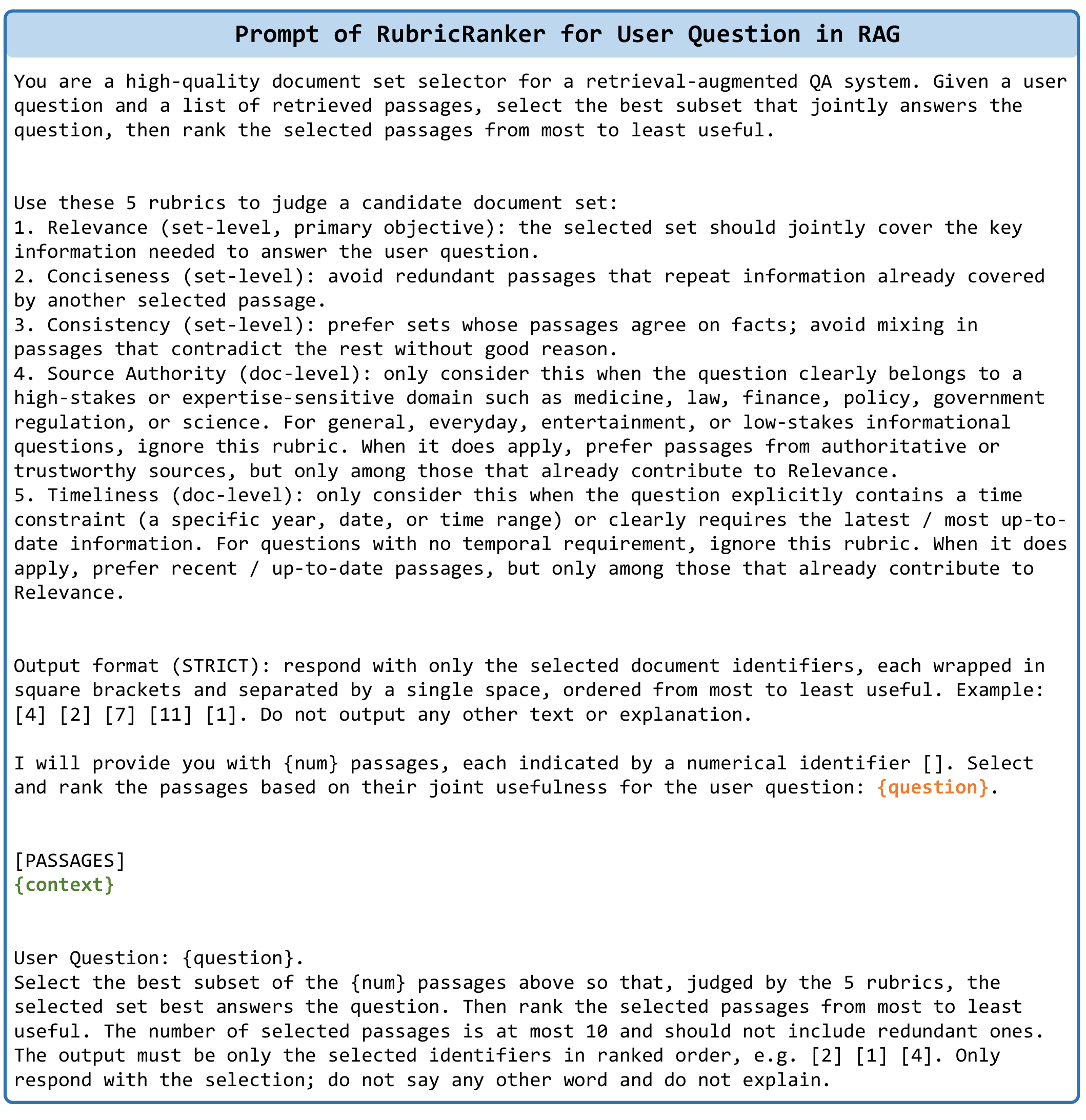}
  \caption{The RubricRanker prompt for a user question in the RAG scenario. The input consists of the user question and candidate documents.}
  \label{fig:rubricrank_prompt_userq}
\end{figure*}

\end{document}